\documentclass{article}
\usepackage{epsfig}
\usepackage{cite}
\def\lapprox{\lower .7ex\hbox{$\;\stackrel{\textstyle <}{\sim}\;$}}
\def\gapprox{\lower .7ex\hbox{$\;\stackrel{\textstyle >}{\sim}\;$}}

\def\e{\epsilon}
\def\d{{\rm d}}
\def\Li{\, {\rm Li}}
\def\S{\, {\rm S}}
\def\sab{s_{12}}
\def\sac{s_{13}}
\def\sbc{s_{23}}

\def\sabc{s_{123}}

\newcommand{\bubbleNLO}[1]{
\mbox{\parbox{2.5cm}{\hspace{0.25cm}
\begin{picture}(2,1)
\thicklines
\put(0.3,0.5){\vector(1,0){0.1}}
\put(0,0.5){\line(1,0){2}}
\put(1,0.5){\circle{1}}
\put(0.25,0.7){\makebox(0,0)[b]{$#1$}}
\end{picture}
}}
\hfill}

\newcommand{\doublebubbleNLO}[1]{
\mbox{\parbox{3.5cm}{\hspace{0.25cm}
\begin{picture}(3,1)
\thicklines
\put(0.3,0.5){\vector(1,0){0.1}}
\put(0,0.5){\line(1,0){0.5}}
\put(2.5,0.5){\line(1,0){0.5}}
\put(1,0.5){\circle{1}}
\put(2,0.5){\circle{1}}
\put(0.25,0.7){\makebox(0,0)[b]{$#1$}}
\end{picture}
}}
\hfill}

\newcommand{\trianglexNLO}[3]{
\mbox{\parbox{3cm}{\hspace{0.25cm}
\begin{picture}(2.5,1.4)
\thicklines
\put(0.3,0.7){\vector(1,0){0.1}}
\put(1.7,0.2){\vector(1,0){0.1}}
\put(1.7,1.2){\vector(1,0){0.1}}
\put(0,0.7){\line(1,0){0.5}}
\put(0.5,1.2){\oval(1,1)[br]}
\put(1,1.2){\line(1,0){1}}
\put(1,0.2){\line(1,0){1}}
\put(1,0.7){\circle{1}}
\put(0.25,0.9){\makebox(0,0)[b]{$#1$}}
\put(2.05,1.2){\makebox(0,0)[l]{$#2$}}
\put(2.05,0.2){\makebox(0,0)[l]{$#3$}}
\end{picture}
}}
\hfill}

\newcommand{\triangleNLO}[3]{
\mbox{\parbox{3cm}{\hspace{0.25cm}
\begin{picture}(2.5,1.4)
\thicklines
\put(0.3,0.7){\vector(1,0){0.1}}
\put(1.7,0.2){\vector(1,0){0.1}}
\put(1.7,1.2){\vector(1,0){0.1}}
\put(0,0.7){\line(1,0){0.5}}
\put(1,1.2){\line(0,-1){1}}
\put(1,1.2){\line(1,0){1}}
\put(1,0.2){\line(1,0){1}}
\put(1,0.7){\circle{1}}
\put(0.25,0.9){\makebox(0,0)[b]{$#1$}}
\put(2.05,1.2){\makebox(0,0)[l]{$#2$}}
\put(2.05,0.2){\makebox(0,0)[l]{$#3$}}
\end{picture}
}}
\hfill}

\newcommand{\trianglebNLO}[3]{
\mbox{\parbox{3cm}{\hspace{0.25cm}
\begin{picture}(2.5,1.4)
\thicklines
\put(0.3,0.7){\vector(1,0){0.1}}
\put(1.9,0.2){\vector(1,0){0.1}}
\put(1.9,1.2){\vector(1,0){0.1}}
\put(0,0.7){\line(1,0){0.5}}
\put(0.5,0.7){\line(2,1){1}}
\put(0.5,0.7){\line(2,-1){1}}
\put(1.5,1.2){\line(0,-1){1}}
\put(1.5,1.2){\line(1,0){0.5}}
\put(1.5,0.2){\line(1,0){0.5}}
\put(1.5,0.2){\line(-1,2){0.4}}
\put(0.25,0.9){\makebox(0,0)[b]{$#1$}}
\put(2.05,1.2){\makebox(0,0)[l]{$#2$}}
\put(2.05,0.2){\makebox(0,0)[l]{$#3$}}
\end{picture}
}}
\hfill}

\newcommand{\boxbubbleaNLO}[4]{
\mbox{\parbox{4cm}{\hspace{0.25cm}
\begin{picture}(3.5,1.4)
\thicklines
\put(0.7,0.2){\vector(-1,0){0.1}}
\put(2.8,0.2){\vector(1,0){0.1}}
\put(2.8,1.2){\vector(1,0){0.1}}
\put(0.8,1.2){\vector(1,0){0.1}}
\put(0.5,0.2){\line(1,0){2.5}}
\put(0.5,1.2){\line(1,0){2.5}}
\put(1,0.2){\line(0,1){1}}
\put(2,0.7){\circle{1}}
\put(0.45,1.2){\makebox(0,0)[r]{$#1$}}
\put(0.45,0.2){\makebox(0,0)[r]{$#2$}}
\put(3.05,1.2){\makebox(0,0)[l]{$#3$}}
\put(3.05,0.2){\makebox(0,0)[l]{$#4$}}
\end{picture}
}} 
\hfill}

\newcommand{\boxbubblebNLO}[4]{
\mbox{\parbox{4cm}{\hspace{0.25cm}
\begin{picture}(3.5,1.4)
\thicklines
\put(0.7,0.2){\vector(-1,0){0.1}}
\put(2.8,0.2){\vector(1,0){0.1}}
\put(2.8,1.2){\vector(1,0){0.1}}
\put(0.8,1.2){\vector(1,0){0.1}}
\put(0.5,0.2){\line(1,0){2.5}}
\put(0.5,1.2){\line(1,0){2.5}}
\put(2.5,0.2){\line(0,1){1}}
\put(1.5,0.7){\circle{1}}
\put(0.45,1.2){\makebox(0,0)[r]{$#1$}}
\put(0.45,0.2){\makebox(0,0)[r]{$#2$}}
\put(3.05,1.2){\makebox(0,0)[l]{$#3$}}
\put(3.05,0.2){\makebox(0,0)[l]{$#4$}}
\end{picture}
}} 
\hfill}

\newcommand{\boxxaNLO}[4]{
\mbox{\parbox{3.5cm}{\hspace{0.25cm}
\begin{picture}(3,1.4)
\thicklines
\put(0.7,0.2){\vector(-1,0){0.1}}
\put(2.3,0.2){\vector(1,0){0.1}}
\put(2.3,1.2){\vector(1,0){0.1}}
\put(0.8,1.2){\vector(1,0){0.1}}
\put(0.5,0.2){\line(1,0){2}}
\put(2,0.2){\line(-1,1){1}}
\put(0.5,1.2){\line(1,0){2}}
\put(1,0.2){\line(0,1){1}}
\put(2,0.2){\line(0,1){1}}
\put(0.45,1.2){\makebox(0,0)[r]{$#1$}}
\put(0.45,0.2){\makebox(0,0)[r]{$#2$}}
\put(2.55,1.2){\makebox(0,0)[l]{$#3$}}
\put(2.55,0.2){\makebox(0,0)[l]{$#4$}}
\end{picture}
}} 
\hfill}

\newcommand{\boxxbNLO}[4]{
\mbox{\parbox{3.5cm}{\hspace{0.25cm}
\begin{picture}(3,1.4)
\thicklines
\put(0.7,0.2){\vector(-1,0){0.1}}
\put(2.3,0.2){\vector(1,0){0.1}}
\put(2.3,1.2){\vector(1,0){0.1}}
\put(0.8,1.2){\vector(1,0){0.1}}
\put(0.5,0.2){\line(1,0){2}}
\put(1,0.2){\line(1,1){1}}
\put(0.5,1.2){\line(1,0){2}}
\put(1,0.2){\line(0,1){1}}
\put(2,0.2){\line(0,1){1}}
\put(0.45,1.2){\makebox(0,0)[r]{$#1$}}
\put(0.45,0.2){\makebox(0,0)[r]{$#2$}}
\put(2.55,1.2){\makebox(0,0)[l]{$#3$}}
\put(2.55,0.2){\makebox(0,0)[l]{$#4$}}
\end{picture}
}} 
\hfill}

\newcommand{\boxxbpNLO}[4]{
\mbox{\parbox{3.5cm}{\hspace{0.25cm}
\begin{picture}(3,1.4)
\thicklines
\put(0.7,0.2){\vector(-1,0){0.1}}
\put(2.3,0.2){\vector(1,0){0.1}}
\put(2.3,1.2){\vector(1,0){0.1}}
\put(0.8,1.2){\vector(1,0){0.1}}
\put(0.5,0.2){\line(1,0){2}}
\put(1.0,0.2){\line(1,2){0.5}}
\put(0.5,1.2){\line(1,0){2}}
\put(1,0.2){\line(0,1){1}}
\put(2,0.2){\line(0,1){1}}
\put(0.45,1.2){\makebox(0,0)[r]{$#1$}}
\put(0.45,0.2){\makebox(0,0)[r]{$#2$}}
\put(2.55,1.2){\makebox(0,0)[l]{$#3$}}
\put(2.55,0.2){\makebox(0,0)[l]{$#4$}}
\end{picture}
}} 
\hfill}

\newcommand{\doubleboxNLO}[4]{
\mbox{\parbox{3.5cm}{\hspace{0.25cm}
\begin{picture}(3,1.4)
\thicklines
\put(0.7,0.2){\vector(-1,0){0.1}}
\put(2.3,0.2){\vector(1,0){0.1}}
\put(2.3,1.2){\vector(1,0){0.1}}
\put(0.8,1.2){\vector(1,0){0.1}}
\put(0.5,0.2){\line(1,0){2}}
\put(1,0.2){\line(0,1){1}}
\put(0.5,1.2){\line(1,0){2}}
\put(1.5,0.2){\line(0,1){1}}
\put(2,0.2){\line(0,1){1}}
\put(0.45,1.2){\makebox(0,0)[r]{$#1$}}
\put(0.45,0.2){\makebox(0,0)[r]{$#2$}}
\put(2.55,1.2){\makebox(0,0)[l]{$#3$}}
\put(2.55,0.2){\makebox(0,0)[l]{$#4$}}
\end{picture}
}}
\hfill}

\newcommand{\doubleboxNLOtwo}[4]{
\mbox{\parbox{3.5cm}{\hspace{0.25cm}
\begin{picture}(3,1.4)
\thicklines
\put(0.7,0.2){\vector(-1,0){0.1}}
\put(2.3,0.2){\vector(1,0){0.1}}
\put(2.3,1.2){\vector(1,0){0.1}}
\put(0.8,1.2){\vector(1,0){0.1}}
\put(0.5,0.2){\line(1,0){2}}
\put(1,0.2){\line(0,1){1}}
\put(0.5,1.2){\line(1,0){2}}
\put(1.5,0.2){\line(0,1){1}}
\put(2,0.2){\line(0,1){1}}
\put(0.45,1.2){\makebox(0,0)[r]{$#1$}}
\put(0.45,0.2){\makebox(0,0)[r]{$#2$}}
\put(2.55,1.2){\makebox(0,0)[l]{$#3$}}
\put(2.55,0.2){\makebox(0,0)[l]{$#4$}}
\put(1.075,1.0){\makebox(0,0)[l]{$_{(2)}$}}
\end{picture}
}}
\hfill}

\newcommand{\boxxbdotNLO}[4]{
\mbox{\parbox{3.5cm}{\hspace{0.25cm}
\begin{picture}(3,1.4)
\thicklines
\put(0.7,0.2){\vector(-1,0){0.1}}
\put(2.3,0.2){\vector(1,0){0.1}}
\put(2.3,1.2){\vector(1,0){0.1}}
\put(0.8,1.2){\vector(1,0){0.1}}
\put(0.5,0.2){\line(1,0){2}}
\put(1,0.2){\line(1,1){1}}
\put(1.5,0.7){\circle*{0.2}}
\put(0.5,1.2){\line(1,0){2}}
\put(1,0.2){\line(0,1){1}}
\put(2,0.2){\line(0,1){1}}
\put(0.45,1.2){\makebox(0,0)[r]{$#1$}}
\put(0.45,0.2){\makebox(0,0)[r]{$#2$}}
\put(2.55,1.2){\makebox(0,0)[l]{$#3$}}
\put(2.55,0.2){\makebox(0,0)[l]{$#4$}}
\end{picture}
}} 
\hfill}

\begin{document}
\unitlength1cm
\begin{titlepage}
\vspace*{-1cm}
\begin{flushright}
TTP00--20\\
hep-ph/0008287\\
August 2000 
\end{flushright}                                
\vskip 3.5cm
\begin{center}
\boldmath
{\Large\bf Two-Loop Master Integrals for $\gamma^* \to 3$ Jets: \\[3mm]
The planar topologies }\unboldmath
\vskip 1.cm
{\large  T.~Gehrmann}$^a$ and {\large E.~Remiddi}$^b$ 
\vskip .7cm
{\it $^a$ Institut f\"ur Theoretische Teilchenphysik,
Universit\"at Karlsruhe, D-76128 Karlsruhe, Germany}
\vskip .4cm
{\it $^b$ Dipartimento di Fisica,
    Universit\`{a} di Bologna and INFN, Sezione di 
    Bologna,  I-40126 Bologna, Italy} 
\end{center}
\vskip 2.6cm

\begin{abstract}
The calculation of the two-loop corrections to the three jet production 
rate and to event shapes in electron-positron annihilation requires 
the computation of a number of up to now unknown two-loop four-point 
master 
integrals with one off-shell and three on-shell legs. In this paper,
we compute those master integrals which correspond to planar
topologies by solving differential equations in the external 
invariants which are fulfilled  by the master integrals. 
We obtain the master integrals 
 as expansions in $\e=(4-d)/2$, where $d$ is the
space-time dimension. The results are expressed in terms of 
newly introduced two-dimensional harmonic polylogarithms, whose
properties are shortly discussed. For all two-dimensional 
harmonic polylogarithms appearing in the divergent parts of the 
integrals, expressions in terms of 
Nielsen's polylogarithms are given. The analytic continuation of 
our results to other kinematical situations is outlined.

\end{abstract}
\vfill
\end{titlepage}                                                                
\newpage

\renewcommand{\theequation}{\mbox{\arabic{section}.\arabic{equation}}}

\section{Introduction}
\setcounter{equation}{0}

The calculation of perturbative next-to-next-to-leading order
corrections to $2\to 2$ scattering or $1 \to 3$ decay processes 
is a yet outstanding task for many precision applications in 
particle physics phenomenology. Progress in this field was 
up to very recently hampered by difficulties in the calculation of the 
virtual two-loop corrections to the corresponding Feynman amplitudes. 

Using dimensional regularisation~\cite{dreg1,dreg2,hv} 
with $d=4-2\e$ dimensions as regulator for
ultraviolet and infrared divergences, 
the large number of different integrals appearing in the 
two-loop Feynman amplitudes for $2\to 2$ scattering or $1\to 3$ decay 
processes 
can be reduced to a small number of master integrals. 
The techniques used in these reduction are 
integration-by-parts identities~\cite{hv,chet1,chet2} 
and Lorentz invariance~\cite{gr}. A computer algorithm for the 
automatic reduction of all two-loop four-point integrals 
was described in~\cite{gr}. 

For two-loop four-point functions with massless internal propagators
and  all legs on-shell, which are 
relevant for example in the next-to-next-to-leading order
calculation of two jet production at hadron colliders, all master
integrals have been calculated over the past
year~\cite{onshell1,onshell2,onshell3,onshell4,onshell5,onshell6}.
To compute the next-to-next-to-leading order corrections to 
observables such as the three jet production rate in electron-positron
annihilation, two plus one jet production in deep inelastic
electron-proton scattering or vector boson plus jet production at hadron 
colliders, one requires a different class of integrals: two-loop
four-point functions with massless internal propagators and 
 one external leg off-shell. Since these 
functions involve one scale more than their on-shell counterparts, one 
expects the corresponding master integrals to be more
complicated, and also to be more numerous. 
First progress towards the computation of these master integrals has
been made very recently by Smirnov, who computed the planar double 
box integral with all propagators raised to unit power~\cite{smirnew}.
Using a Mellin-Barnes contour integral technique, Smirnov 
obtained an analytic expression for the divergent parts and a
one-dimensional integral representation for the finite term. 
Complementary work on the purely numerical evaluation of this 
type of master integrals has been presented recently by Binoth and 
Heinrich~\cite{num}.

In this paper, we compute all master integrals appearing in the
reduction of planar two-loop
four-point functions with one external leg off-shell.
The method employed here is substantially different from the 
techniques used in~\cite{smirnew,num}: for all master integrals under
consideration, we derived inhomogeneous
differential equations~\cite{gr} in the 
external invariants. The master integrals are then determined 
by solving these equations in terms of newly introduced two-dimensional 
harmonic polylogarithms and subsequent matching of the solution to the 
boundary condition. 

In Section~\ref{sec:de}, we briefly review the use of the differential 
equation approach to the computation of master integrals. To solve 
the differential equations for two-loop four-point master integrals
with one off-shell leg, we employ an ansatz which is 
the product of a rational function of the kinematic invariants times 
the sum of two-dimensional harmonic polylogarithms (2dHPL). We explain how the 
unknown parameters in the ansatz are determined from the 
differential equation and its boundary condition. To illustrate how the 
method works in practice, we outline the calculation of two master
integrals in Section~\ref{sec:example}.

The master integrals computed using this method are listed 
in Section~\ref{sec:mi}, 
which also contains a reference list of simpler master
integrals for planar two- and there-point functions, which appear 
in the reduction of the four-point functions considered here.
Section~\ref{sec:conc} contains concluding remarks and an outlook on
potential applications and open problems. 

We enclose an Appendix which summarises the properties of the harmonic 
polylogarithms (HPL)~\cite{hpl} and extends the underlying formalism 
towards the newly introduced two-dimensional harmonic polylogarithms 
(2dHPL). In the Appendix, we also tabulate relations between the HPL/2dHPL 
 and the more commonly known Nielsen's generalised 
polylogarithms~\cite{nielsen,bit}. 
We then give the expression of all (ordinary and two-dimensional) 
harmonic polylogarithms appearing in the divergent parts of the master 
integrals computed in this paper in terms of generalised polylogarithms 
of suitable arguments. 
As a consequence, the harmonic polylogarithms of weight 4 appearing 
in the finite parts of the master integrals can all be represented as 
one-dimensional integrals over ordinary Nielsen's polylogarithms up to 
weight three. Their precise numerical evaluation can be obtained by a 
simple extension of the techniques of~\cite{bit}.

\section{Differential Equations for Master Integrals}
\label{sec:de}
\setcounter{equation}{0}

The use of differential equations for the computation of master integrals 
was first suggested in~\cite{kotikov} as means to relate integrals 
with massive internal propagators to their massless counterparts. The
method was developed in detail in~\cite{remiddi}, where it was also 
extended towards differential equations in the external invariants. 
As a first application of this method, the two-loop sunrise diagram 
with arbitrary internal masses was studied in~\cite{appl}. We 
have developed the differential equation formalism for
two-loop four-point functions with massless internal propagators,
three external legs on-shell and
one external leg off-shell in~\cite{gr}.
We derived an algorithm for the automatic reduction 
by means of computer algebra (using FORM~\cite{form} and Maple~\cite{maple})
of any
two-loop four-point integral to a small set of master integrals, for which we 
derived differential equations in the external invariants. For all 
master integrals with up to five internal propagators, we 
solved the differential equations in a closed form for arbitrary 
number of space-time dimensions $d$. 
In the present work, we elaborate on these results by deriving 
Laurent expansions around $\e=0$ for all master integrals derived 
in~\cite{gr} as well as all remaining planar master integrals with 
six or seven propagators. 

Four-point functions depend on three linearly independent momenta:
$p_1$, $p_2$ and $p_3$. In our calculation, we take all these momenta 
on-shell ($p_i^2=0$), 
while the fourth momentum $p_{123}=(p_1+p_2+p_3)$ is taken
off-shell. The kinematics of the four-point function is then 
fully described by specifying the values of the three Lorentz 
invariants $\sab = (p_1+p_2)^2$, $\sac = (p_1+p_3)^2$ and $\sbc =
(p_2+p_3)^2$. 

Expressing the system of differential equations obtained in~\cite{gr} for 
any master integral in the variables 
$\sabc = \sab + \sac + \sbc$, $y=\sac/\sabc$ and $z=\sbc/\sabc$, we
obtain a homogeneous equation in $\sabc$, and inhomogeneous equations 
in $y$ and $z$. Since $\sabc$ is the only quantity carrying a mass 
dimension, 
the corresponding differential equation is nothing but the rescaling 
relation obtained by investigating the behaviour of the master integral 
under a rescaling of all external momenta by a constant factor. 

Some of the master integrals under consideration do not depend explicitly
on all three invariants, but only on certain combinations of them, thus
being one-scale or two-scale integrals. The one-scale integrals can not
be determined using the differential equation method, since the 
only non-trivial differential equation fulfilled by them is the 
homogeneous rescaling relation. These integrals are relatively simple
and can all be computed using Feynman parameters~\cite{kl1,kl2}. 
The two-scale integrals
fulfil, besides the rescaling relation, one inhomogeneous differential 
equation, which can be employed for their computation. These integrals  
have all been computed in~\cite{daw1,daw2,glover,gr}, 
where they are given in a 
closed form for arbitrary $\e$.

The solution of differential equations for two-loop four-point functions 
with one off-shell leg in terms of hypergeometric functions, was
discussed in detail in~\cite{gr}. Although this method yields at first 
sight very compact results for arbitrary $\e$, it is not very well
suited for practical applications, where an expansion around 
$\e=0$ is needed. The major problem with this approach
is the appearance of generalised hypergeometric functions
which can be expressed only in terms of multidimensional integrals.
The evaluation of these multiple integrals to the required order of 
the $\e$-expansion is a problem comparable to the direct evaluation of
the master integral from its Feynman parameter representation. 
To solve the differential equations for the three-scale integrals, we 
choose therefore an approach different from the one pursued in~\cite{gr}. 

The procedure employed here is as follows. In the
$y$ and $z$  differential equations 
for the master integral under consideration, the
coefficient of the homogeneous term as well as the full inhomogeneous
term (coefficients and subtopologies) are 
expanded as a series in $\e$.
For the master integral, we make an ansatz which has the form
\begin{equation}
\sum_i{\cal R}_i(y,z;\sabc,\e){\cal H}_i(y,z;\e)\; ,
\label{eq:ansatz}
\end{equation}
where the prefactor
${\cal R}_i(y,z;\sabc,\e)$ is a rational function of $y$ and $z$,
which is multiplied with an overall normalisation factor to account for
the correct dimension in $\sabc$, while ${\cal H}_i(y,z;\e)$ is a Laurent
series in $\e$. The coefficients of its $\e$-expansion are then written
as the sum of 
2dHPL up to a weight 
determined by the order of the series:
\begin{equation}
{\cal H}_i(y,z;\e) = \frac{\e^p}{\e^4} \sum_{n=0}^{4} \e^n
\left[ T_n(z) + 
\sum_{j=1}^n \sum_{\vec{m}_j\in V_{j}(z)} T_{n,\vec{m}_j}(z) H(\vec{m}_j;y)
\right] \; , 
\label{eq:calH}
\end{equation} 
where the $ H(\vec{m}_j;y) $ are 2dHPL and 
$ T_n(z)$, $T_{n,\vec{m}_j}(z) $ are the as yet unknown coefficients. 
This form is motivated by the expectation that 
two-loop four-point functions can diverge up to $1/\e^4$, as obtained by 
considering soft and collinear limits of internal
propagators~\cite{catani}; 
$p$ is an --~a priori unknown~-- integer
number, accounting for the fact that some of the  
master integrals might have a lower superficial degree of divergence. 
$V_{j}(z)$ is the 
set of all possible indices for 2dHPL of weight $j$ ($j$-dimensional 
vectors made from the components 
$(0,1,1-z,z)$). 
The $T_n(z)$ and $T_{n,\vec{m}}(z)$ are unknown
coefficients, which can contain in turn 
ordinary one-dimensional HPL depending on $z$. With the ansatz 
(\ref{eq:calH}), the leading $1/\e^4$-singularity corresponds to 
$p=0$ and $n=0$.

The rational functions ${\cal R}_i(y,z;\sabc,\e)$ can be determined 
from the homogeneous part of the differential equations in $y$ and $z$
by inserting only the constant $n=0$ term of ${\cal H}_i(y,z;\e)$.

It turns out that for all topologies which contain 
a single master integral, only one ${\cal R}(y,z;\sabc,\e)$, and 
consequently also only one  ${\cal H}(y,z;\e)$, are
present, such that the sum in (\ref{eq:ansatz}) is not needed. 
For both topologies containing two master integrals, only one
${\cal R}(y,z;\sabc,\e)$ and ${\cal H}(y,z;\e)$
per master integral are sufficient, provided
that the basis of first and second master integral is chosen appropriately,
as explained later in this Section. 

Having determined the prefactor
${\cal R}(y,z;\sabc,\e)$ for the master integral
under consideration, we rewrite the $y$ differential 
equation for the master integral into a $y$ differential equation for 
${\cal H}(y,z;\e)$. From the $\e$-expansion of the inhomogeneous term,
one can read off the value of $p$, required to match the 
order of the 
Laurent series of the master integral to the Laurent series 
of the inhomogeneous term. 

By construction of the 2dHPL, their derivative with respect to 
$y$ is straightforward (see Appendix~\ref{ap:two}):
\begin{eqnarray}
\frac{\d}{\d y} H(m;y) &=& f(m;y)\; , \nonumber \\
\frac{\d}{\d y} H(m,\vec{m}_{j-1};y) &=& f(m;y) H( \vec{m}_{j-1};y)\;,
\end{eqnarray}
where the $f(m;y)$ are nothing but the $y$-dependent denominators 
present in the differential equation. 
Inserting the right hand side of (\ref{eq:calH}) into the 
$y$ differential equation for  ${\cal H}(y,z;\e)$ and differentiating 
according to the above rules, the differential equation becomes a
purely algebraic equation. 
Grouping the different inverse powers of 
$y$, $(1-y)$, $(1-y-z)$ and $(y+z)$
and using the linear independence of the $H(\vec{m}_j;y)$, this 
algebraic equation can be
translated into a linear system of equations for the coefficients 
$T_{n,\vec{m}_j}(z)$. The resulting system of typically several hundreds of 
equations can be solved for the $T_{n,\vec{m}_j}(z)$ using computer
algebra (here we employ the same algorithm as used already in~\cite{gr}
for solving large numbers of integration-by-parts and 
Lorentz-invariance identities). 

The $T_{n}(z)$, which do not multiply any $y$-dependent 
function in the ansatz and can therefore not be determined in the above 
way from the differential equation, correspond to the boundary condition. 
They are determined by exploiting the 
fact that all planar master integrals as well as their derivatives 
are analytic in the whole 
kinematical plane with the exception of the two branch points 
at $y=0$ and $z=0$. As a consequence, any factor $(1-y)$, $(1-y-z)$ or
$(y+z)$ appearing in the denominator of the
homogeneous term of the differential equations 
for a master integral can be used to determine the boundary condition in 
$y=1$, $y=(1-z)$ or $y=-z$ respectively: multiplying the 
differential equation with this factor and taking the limit where the 
factor vanishes, one obtains a linear relation between the master integral 
in this special kinematical point
and its subtopologies, no longer involving any derivatives.  

Some special care has to be taken in the case of topologies
with two master integrals. Our choice of basis for the master integrals 
in these cases is as follows.
For the five-propagator (diagonal box) 
master integrals, we use the basis already
employed in~\cite{gr}:
\begin{eqnarray}
\boxxbNLO{p_{123}}{p_1}{p_2}{p_3} &=& \int \frac{\d^d
k}{(2\pi)^d}\frac{\d^d l}{(2\pi)^d}  \frac{1}{k^2 (k-p_{123})^2 
(k-l-p_2)^2 l^2 (l-p_3)^2}, 
\label{eq:fivea}\\
\boxxbdotNLO{p_{123}}{p_1}{p_2}{p_3} &=&\int \frac{\d^d
k}{(2\pi)^d}\frac{\d^d l}{(2\pi)^d}  \frac{1}{k^2 (k-p_{123})^2 
(k-l-p_2)^4 l^2 (l-p_3)^2}.\label{eq:fiveb}
\end{eqnarray}
For the seven-propagator double box integrals, we use:
\begin{eqnarray}
\doubleboxNLO{p_{123}}{p_1}{p_2}{p_3} &=& \int \frac{\d^d
k}{(2\pi)^d}\frac{\d^d l}{(2\pi)^d}  \frac{1}{k^2 (k-p_{23})^2
(k-p_{123})^2 (k-l)^2 l^2 (l-p_2)^2 (l-p_{23})^2 },
\label{eq:sevena}
\\ 
\doubleboxNLOtwo{p_{123}}{p_1}{p_2}{p_3} &=& \int \frac{\d^d
k}{(2\pi)^d}\frac{\d^d l}{(2\pi)^d}  \frac{(k-p_2)^2}{k^2 (k-p_{23})^2
(k-p_{123})^2 (k-l)^2 l^2 (l-p_2)^2 (l-p_{23})^2 }.
\label{eq:sevenb}
\end{eqnarray}
This choice of basis for the double box integrals has been 
first suggested by Anastasiou, Tausk and Tejeda-Yeomans
in~\cite{onshell6} for the double box integrals with all legs on-shell. 
We find it to be convenient in the off-shell case as well. Apart from 
yielding a compact expression for the second master integral, 
this choice also circumvents problems with subleading 
terms~\cite{onshell7} appearing 
in the reduction of tensor integrals.

For both topologies, one obtains coupled sets of differential 
equations for the two master integrals. If only the $n=0$ term of 
${\cal H}_i(y,z;\e)$ is retained, the equations decouple, thus enabling
the determination of the ${\cal R}_i(y,z;\sabc,\e)$ for each of the 
two master integrals. The boundary conditions for both
master integrals are determined as above by using analyticity
properties in the kinematical variables.   
Boundary conditions in two kinematical points 
are needed to fully constrain both master
integrals. 
In the case of the seven
propagator double box integrals (\ref{eq:sevena},\ref{eq:sevenb}), 
only one of those can be obtained in $y=(1-z)$, no other $y$-dependent 
denominator corresponding to an analytic point 
is present in the homogeneous term. To find the second 
boundary condition, we first determine the $z$-dependence of the
$T_{n}(z)$
by investigating their $z$ differential equations, obtained from 
the $y$ and $z$ differential equations of the master integrals in $y=(1-z)$. 
The remaining constant terms in $T_{n}(z)$ are then obtained by using
the fact that the master integrals are regular in $z=1$. We would like 
to point out that this procedure relies crucially on the possibility
to vary
$y$ and $z$ independently, which is not the case for the 
on-shell double box integral, 
where the kinematical properties of the master integrals 
are insufficient~\cite{onshell5} to determine both boundary
conditions from the differential equation alone.

Following the steps described in this section, we have derived the 
$\e$-expansions of all planar four-point master integrals with 
one off-shell leg. It is obvious that this calculation had to follow 
a bottom-up approach starting from master integrals with few different
denominators, which subsequently appeared in the inhomogeneous term of
the differential equations for master integrals with more 
different denominators. 

As a check on all three-scale integrals obtained with this method, the 
results were inserted into the corresponding $z$ differential equation.
After transforming the $H(\vec{m}_j\in V_{j}(z);y)$ into 
$H(\vec{m}_j\in V_{j}(y);z)$, the $z$ derivatives can be carried out
straightforwardly, thus providing a check on our results.

In the following Section, we illustrate
on the example of the master
integrals (\ref{eq:fivea},\ref{eq:fiveb}) how our method is applied in
practice. 
In Section~\ref{sec:mi}, we list our results for all 
master integrals 
appearing in the reduction of any planar two-loop four-point integral 
with one off-shell leg. 

\section{Computing a Master Integral from its Differential 
Equation}
\label{sec:example}
\setcounter{equation}{0}

The diagonal box master integrals (\ref{eq:fivea},\ref{eq:fiveb})
are symmetric under the interchange of $p_1 \leftrightarrow p_2$, 
and consequently symmetric under $y\leftrightarrow z$. The coupled 
set of $y$ 
differential equations for them reads:
\begin{eqnarray}
\frac{\partial}{\partial y} \boxxbNLO{p_{123}}{p_1}{p_2}{p_3} &=& 
\left[-\frac{1}{y+z} + (d-4) \left(\frac{1}{y} - \frac{1}{4}
\frac{1}{1-y-z} -2\frac{1}{y+z} \right)\right]
\boxxbNLO{p_{123}}{p_1}{p_2}{p_3} \nonumber \\
&& 
+\frac{1}{4} \frac{yz}{(1-y-z)(y+z)} 
  \boxxbdotNLO{p_{123}}{p_1}{p_2}{p_3} \nonumber \\
&& 
+ \frac{(d-3)(3d-10)}{4(d-4)} \frac{-4 + 4z +
5y}{y(y+z)(1-y-z)} \triangleNLO{p_{123}}{p_{13}}{p_2} \nonumber\\
&& 
- \frac{(d-3)(3d-10)}{4(d-4)} \frac{-4 + 4z +
3y}{y(y+z)(1-y-z)} \triangleNLO{p_{123}}{p_{23}}{p_1} \nonumber\\
&&
 - \frac{(d-3)(3d-8)(3d-10)}{2(d-4)^2} \frac{1}{y(y+z)(1-y-z)}
\bubbleNLO{p_{13}} \nonumber\\
&&
- \frac{(d-3)(3d-8)(3d-10)}{2(d-4)^2} \frac{1}{z(y+z)(1-y-z)}
\bubbleNLO{p_{23}} \; ,\label{eq:difexa}\\
\frac{\partial}{\partial y} \boxxbdotNLO{p_{123}}{p_1}{p_2}{p_3} &=& 
\left[-\frac{1}{y} + (d-4) \left( \frac{1}{y} + \frac{3}{4}
\frac{1}{1-y-z}\right) \right] \boxxbdotNLO{p_{123}}{p_1}{p_2}{p_3} 
\nonumber \\
&& 
- \frac{3(d-4)^2}{4} \frac{y+z}{yz(1-y-z)}
\boxxbNLO{p_{123}}{p_1}{p_2}{p_3} \nonumber \\
&&
+ \frac{3(d-3)(3d-10)}{4} \frac{-1+y+2z}{y(1-y)z(1-y-z)} 
\triangleNLO{p_{123}}{p_{13}}{p_2}\nonumber\\
&& 
+ \frac{3(d-3)(3d-10)}{4} \frac{1}{yz(1-y-z)} 
\triangleNLO{p_{123}}{p_{23}}{p_1}
 \nonumber\\
&& 
-\frac{3(d-3)(3d-8)(3d-10)}{2(d-4)} \frac{1}{y^2(1-y)(1-y-z)} 
\bubbleNLO{p_{13}} \nonumber\\
&&
-\frac{3(d-3)(3d-8)(3d-10)}{2(d-4)} \frac{1}{yz^2(1-y-z)} 
\bubbleNLO{p_{23}} \; .\label{eq:difexb}
\end{eqnarray}
From the powers of $(d-4)$ multiplying the inhomogeneous terms, it can
be read off that the second master integral (\ref{eq:fiveb}) is suppressed by 
one power of $(d-4)$ with respect to the first 
master integral (\ref{eq:fivea}). 
Expanding the 
homogeneous term of the differential equations and retaining only the
leading contribution, the homogeneous differential equations decouple:
\begin{eqnarray}
\frac{\partial}{\partial y} \boxxbNLO{p_{123}}{p_1}{p_2}{p_3} &=& 
-\frac{1}{y+z} \boxxbNLO{p_{123}}{p_1}{p_2}{p_3} \; , \\
\frac{\partial}{\partial y} \boxxbdotNLO{p_{123}}{p_1}{p_2}{p_3} &=& 
-\frac{1}{y}\boxxbdotNLO{p_{123}}{p_1}{p_2}{p_3}\; .
\end{eqnarray}
The homogeneous $z$ differential equations follow from the above by
$y\leftrightarrow z$, using the $y\leftrightarrow z$ interchange 
symmetry of the integrals.
As a result, one finds the prefactors 
for the first master integral (\ref{eq:fivea}): 
${\cal R}(y,z;\sabc,\e) \sim 1/(y+z)$ and for the second master integral
(\ref{eq:fiveb}): ${\cal R}(y,z;\sabc,\e) \sim 1/(yz)$. 

Separating off these prefactors, we can transform
(\ref{eq:difexa},\ref{eq:difexb}) into a set of coupled differential 
equations for the two ${\cal H}(y,z;\e)$, corresponding to 
both master integrals. Inserting the ansatz (\ref{eq:calH}) for the
${\cal H}(y,z;\e)$, we can then determine by comparison with the 
$\e$-expansion of the inhomogeneous term that 
$p=0$ for (\ref{eq:fivea}) and 
$p=1$ for (\ref{eq:fiveb}). 
The coefficients $T_{n,\vec{m}_j}(z)$ appearing in the ansatz 
are then determined as outlined in the previous Section. 

To determine the boundary terms $T_n(z)$ for both master integrals, 
we first investigate the differential equations
(\ref{eq:difexa},\ref{eq:difexb}) in the point $y=1-z$, where both
master integrals are analytic. Multiplying both equations with $(1-y-z)$ 
and taking $y=1-z$, the left hand sides of the equations vanish.
From (\ref{eq:difexa}) we obtain,
\begin{equation}
0 = -\frac{d-4}{4} \boxxbNLO{p_{123}}{p_1}{p_2}{p_3} \Bigg|_{y=1-z}
+ \frac{z(1-z)}{4} \boxxbdotNLO{p_{123}}{p_1}{p_2}{p_3} \Bigg|_{y=1-z}
+ (\mbox{Subtopologies})\; ,
\label{eq:limitfive}
\end{equation}
while (\ref{eq:difexb}) yields a multiple of this. The point $y=(1-z)$
can therefore be used only to determine a linear combination of boundary 
terms from both master integrals. A second boundary condition is
obtained by multiplying (\ref{eq:difexa}) with $(y+z)$ and 
subsequently taking $y=-z$,
which fixes the boundary terms for the first master integral
(\ref{eq:fivea}). The boundary terms for the second master integral 
(\ref{eq:fiveb}) follow then from (\ref{eq:limitfive}), thus completing
the computation of both master integrals. The results are 
given in 
the following section, together with the results for the other master
integrals.

\section{Master Integrals}
\label{sec:mi}
\setcounter{equation}{0}

In this Section, we tabulate all master integrals relevant for the
computation of planar two-loop four point functions with one off-shell leg. 
We classify the integrals according to the number of different 
kinematical scales on 
which they depend into one-scale integrals, two-scale integrals and 
three-scale integrals. 

The common normalisation factor of all master integrals is 
\begin{equation}
S_\e = \left[(4\pi)^\e \frac{ \Gamma (1+\e)
    \Gamma^2 (1-\e)}{ \Gamma (1-2\e)} \right] \; .
\end{equation}

\subsection{One-scale Integrals}
\label{sec:mi1}

The one-scale two-loop integrals with massless internal propagators 
were computed a long time ago in the context of the two-loop QCD corrections 
to the photon-quark-antiquark vertex~\cite{kl1,kl2}. We list these 
integrals here only for completeness:
\begin{eqnarray}
\doublebubbleNLO{p_{12}} & = & \left(\frac{S_\e}{16 \pi^2}\right)^2\,
 \left( -s_{12} \right)^{-2\e}\, \frac{-1}{\e^2(1-2\e)^2}\;,\\
\bubbleNLO{p_{12}} & = & \left(\frac{S_\e}{16 \pi^2}\right)^2\,
 \left( -s_{12} \right)^{1-2\e}\,
\bigg[ -\frac{1}{4\e} - \frac{13}{8} -\frac{115}{16}\e 
-\left( \frac{865}{32} - \frac{3}{2}\zeta_3\right)\e^2 \nonumber \\
&&       - \left(\frac{5971}{64} - \frac{39}{4}\zeta_3 
-\frac{\pi^4}{40}\right) \e^3 + {\cal
O}(\e^4)\bigg]\;,\\
\triangleNLO{p_{12}}{p_1}{p_2} & = & \left(\frac{S_\e}{16 \pi^2}\right)^2\,
 \left( -s_{12} \right)^{-2\e}\,
\bigg[ -\frac{1}{2\e^2} - \frac{5}{2\e} - \left(  \frac{19}{2} 
+ \frac{\pi^2}{6} \right) - \left(\frac{65}{2}
+ \frac{5\pi^2}{6} - 2\zeta_3 
\right)\e\nonumber \\
&&
- \left( \frac{211}{2} + \frac{19\pi^2}{6} -10\zeta_3 \right)\e^2 
+ {\cal O}(\e^3) \bigg]\;.
\end{eqnarray}

\subsection{Two-scale Integrals}
\label{sec:mi2}

The relevant two-scale integrals, corresponding to three-point functions
with two off-shell legs,  fulfil one inhomogeneous differential equation 
in the ratio of the two external invariants. Following the strategy 
outlined in Section~\ref{sec:de}, we computed these master integrals by 
employing a product ansatz of the form (\ref{eq:ansatz}), with ${\cal H}$ 
containing only one-dimensional harmonic polylogarithms. 

Results for all two-scale integrals 
existed already in the literature in a closed form for 
arbitrary $\e$ in terms of ordinary hypergeometric 
functions~\cite{daw1,daw2,glover,gr}. 
We checked that our results, obtained from 
solving the differential equations for these integrals, 
agree with the $\e$-expansions of the hypergeometric functions 
quoted in~\cite{daw1,daw2,glover,gr}. 

Our results for the two-scale integrals read:
\begin{equation}
\triangleNLO{p_{123}}{p_{12}}{p_3}  =  \left(\frac{S_\e}{16
\pi^2}\right)^2\, \left( -\sabc \right)^{-2\e}
\sum_{i=-2}^2 \frac{g_{4.1,i}
\left(\frac{\sab}{\sabc}\right)}{\e^i}\; + 
{\cal O}(\e^3) ,
\end{equation}
with:
\begin{eqnarray}
g_{4.1,2}(x) &=& -\frac{1}{2} \ , \\
g_{4.1,1}(x) &=& -\frac{5}{2} \ , \\
g_{4.1,0}(x) &=& -\frac{19}{2} - H(1,0;x) - \frac{\pi^2}{6} \ , \\
g_{4.1,-1}(x) &=&           
          - \frac{65}{2}
          - 5H(1,0;x)
          + 2H(1,0,0;x)
          - H(1,1,0;x)
          + 2\zeta_3
          + \frac{\pi^2}{6} \left[ -5 -H(1;x)\right] \ ,  \\
g_{4.1,-2}(x) &=&        
          - \frac{211}{2}
          - 19H(1,0;x)
          + 10H(1,0,0;x)
          - 4H(1,0,0,0;x)
          - 5H(1,1,0;x)
          + 2H(1,1,0,0;x)\nonumber \\
&&
          - H(1,1,1,0;x)
          + \zeta_3\left[ 10 - H(1;x) \right]
          + \frac{\pi^2}{6} \left[ -19 - 5H(1;x) - H(1,1;x) \right] \ .
\end{eqnarray}

\begin{equation}
\trianglexNLO{p_{123}}{p_3}{p_{12}}  =  \left(\frac{S_\e}{16
\pi^2}\right)^2\, \left( -\sabc \right)^{-2\e}
\sum_{i=-2}^2 \frac{g_{4.2,i}
\left(\frac{\sab}{\sabc}\right)}{\e^i}\; + 
{\cal O}(\e^3) ,
\end{equation}
with:
\begin{eqnarray}
g_{4.2,2}(x) &=& -\frac{1}{2} \ , \\
g_{4.2,1}(x) &=& -\frac{5}{2} +H(0;x)  \ , \\
g_{4.2,0}(x) &=& 
          -\frac{19}{2}           
          + 5H(0;x)
          - H(0,0;x)
          + H(1,0;x) 
          + \frac{\pi^2}{6} \ , \\
g_{4.2,-1}(x) &=& 
          - \frac{65}{2}
          + 19H(0;x)
          - 5H(0,0;x)
          + H(0,0,0;x)
          - H(0,1,0;x)
          + 5H(1,0;x)
          - H(1,0,0;x)\nonumber \\
&&
          + H(1,1,0;x)
          + \zeta_3 
          + \frac{\pi^2}{6} \left[+ 5- H(0;x)+ H(1;x)\right]  \ , \\
g_{4.2,-2}(x) &=& 
          - \frac{211}{2}
          + 65H(0;x)
          - 19H(0,0;x)
          + 5H(0,0,0;x)
          - H(0,0,0,0;x)
          + H(0,0,1,0;x)\nonumber \\
&&
          - 5H(0,1,0;x)
          + H(0,1,0,0;x)
          - H(0,1,1,0;x)
          + 19H(1,0;x)
          - 5H(1,0,0;x)\nonumber \\
&&
          + H(1,0,0,0;x)
          - H(1,0,1,0;x)
          + 5H(1,1,0;x)
          - H(1,1,0,0;x)
          + H(1,1,1,0;x) 
          + \frac{3\pi^4}{40} \nonumber \\
&&
          + \zeta_3 \left[ + 5- 4H(0;x)- 2H(1;x)\right]
          + \frac{\pi^2}{6} \big[         
          + 19
          - 5H(0;x)
          + H(0,0;x)
          - H(0,1;x)\nonumber \\
&&
          + 5H(1;x)
          - H(1,0;x)
          + H(1,1;x) \big] \ .
\end{eqnarray}

\begin{equation}
\trianglebNLO{p_{123}}{p_{12}}{p_3}  =  \left(\frac{S_\e}{16
\pi^2}\right)^2\, \frac{\left( -\sabc \right)^{-2\e}}{\sabc-\sab}
\sum_{i=0}^4 \frac{g_{5.1,i}
\left(\frac{\sab}{\sabc}\right)}{\e^i}\; + 
{\cal O}(\e) ,
\end{equation}
with:
\begin{eqnarray}
g_{5.1,4}(x) &=& g_{5.1,3}(x) = g_{5.1,2}(x) = g_{5.1,1}(x) = 0  \ , \\
g_{5.1,0}(x) &=& 
          - H(0,0,1,0;x)
          + H(0,1,0,0;x)
          - \frac{\pi^2}{6}  H(0,0;x)
          + 3 \zeta_3 H(0;x)
          - \frac{\pi^4}{15} \ .
\end{eqnarray}

\subsection{Three-scale Integrals}
\label{sec:mi3}
It was outlined in Section~\ref{sec:de} that 
each three-scale master integral fulfils two inhomogeneous differential 
equations. For the computation of the master integral, it is sufficient 
to solve one of the differential equations, the second one is then used 
as an independent check on the result. Following the steps discussed in 
 Section~\ref{sec:de} we have computed all three-scale master integrals 
corresponding to planar topologies. 

The two different
bubble insertions into the one-loop box integral with one off-shell leg 
were first considered in~\cite{glover}, where they were computed using the 
negative dimension approach. We confirmed~\cite{gr} 
these results 
in the  differential equation approach. Both~\cite{glover} and~\cite{gr}
quoted a result in terms of generalised hypergeometric functions, 
valid for arbitrary $\e$. As explained above, there is no 
straightforward procedure to expand these generalised
hypergeometric functions in a Laurent series in $\e$.
Performing the Laurent series in $\e$ on the differential equations
 yields the 
$\e$-expansion of the master integrals without the need  to expand 
generalised hypergeometric functions.
We obtain for
bubble insertions into the one-loop box integral:
\begin{equation}
\boxbubbleaNLO{p_{123}}{p_1}{p_2}{p_3} = \left(\frac{S_\e}{16
\pi^2}\right)^2\, \frac{\left( -\sabc \right)^{-2\e}}
{\sab+\sac} \sum_{i=-1}^3 \frac{f_{5.1,i}
\left(\frac{\sac}{\sabc},\frac{\sbc}{\sabc}\right)}{\e^i}\; + 
{\cal O}(\e^2) ,
\end{equation}
with:
\begin{eqnarray}
f_{5.1,3}(y,z) &=& 0 \ , \\
f_{5.1,2}(y,z) &=& - H(0;z) \ , \\
f_{5.1,1}(y,z) &=&            
          + H(0;y)H(0;z)
          - 2 H(0;z)
          + 2 H(0,0;z)
          + H(1,0;z) + \frac{\pi^2}{6} \ ,  \\
f_{5.1,0}(y,z) &=&            
          + 2H(0;y)H(0;z)
          - 2H(0;y)H(1,0;z)
          - 4H(0;z)
          - H(0;z)H(1-z,0;y)\nonumber \\
&&
          - 2H(0,0;y)H(0;z)
          + 4H(0,0;z)
          - 2H(0,0;z)H(0;y)
          - 4H(0,0,0;z)
          - H(0,1,0;y)\nonumber \\
&&
          - 2H(0,1,0;z)
          + 2H(1,0;z)
          - H(1,0;z)H(1-z;y)
          - 2H(1,0,0;z)
          - 2H(1,1,0;z)\nonumber \\
&&
          - H(1-z,1,0;y) 
    + \frac{\pi^2}{6} \left[           
          + 2
          - 2H(0;y)
          - 3H(0;z)
          - 2H(1;z)
          - H(1-z;y) \right] \ ,  \\
f_{5.1,-1}(y,z) &=&            
          + 4H(0;y)H(0;z)
          - 4H(0;y)H(1,0;z)
          + 4H(0;y)H(1,0,0;z)
          + 4H(0;y)H(1,1,0;z)\nonumber \\
&&
          - 8H(0;z)
          - 2H(0;z)H(1-z,0;y)
          + 2H(0;z)H(1-z,0,0;y)\nonumber \\
&&
          + H(0;z)H(1-z,1-z,0;y)
          - 4H(0,0;y)H(0;z)
          + 4H(0,0;y)H(0,0;z)\nonumber \\
&&
          + 4H(0,0;y)H(1,0;z)
          + 8H(0,0;z)
          - 4H(0,0;z)H(0;y)
          + 2H(0,0;z)H(1-z,0;y)\nonumber \\
&&
          + 4H(0,0,0;y)H(0;z)
          - 8H(0,0,0;z)
          + 4H(0,0,0;z)H(0;y)
          + 8H(0,0,0,0;z)\nonumber \\
&&
          + 2H(0,0,1,0;y)
          + 4H(0,0,1,0;z)
          - 2H(0,1,0;y)
          - 4H(0,1,0;z)\nonumber \\
&&
          + 4H(0,1,0;z)H(0;y)
          + 2H(0,1,0;z)H(1-z;y)
          + 2H(0,1,0,0;y)
          + 4H(0,1,0,0;z)\nonumber \\
&&
          - H(0,1,1,0;y)
          + 4H(0,1,1,0;z)
          + 2H(0,1-z;y)H(1,0;z)\nonumber \\
&&
          + 2H(0,1-z,0;y)H(0;z)
          + 2H(0,1-z,1,0;y)
          + 4H(1,0;z)
          - 2H(1,0;z)H(1-z;y)\nonumber \\
&&
          + 2H(1,0;z)H(1-z,0;y)
          + H(1,0;z)H(1-z,1-z;y)
          - 4H(1,0,0;z)\nonumber \\
&&
          + 2H(1,0,0;z)H(1-z;y)
          + 4H(1,0,0,0;z)
          + 4H(1,0,1,0;z)
          - 4H(1,1,0;z)\nonumber \\
&&
          + 2H(1,1,0;z)H(1-z;y)
          + 4H(1,1,0,0;z)
          + 4H(1,1,1,0;z)
          + H(1-z,0,1,0;y)\nonumber \\
&&
          - 2H(1-z,1,0;y)
          + 2H(1-z,1,0,0;y)
          - H(1-z,1,1,0;y)
          + H(1-z,1-z,1,0;y)\nonumber \\
&&
          +\frac{7\pi^4}{90}
          +5\zeta_3 H(0;z) \nonumber \\
&&
   + \frac{\pi^2}{6} \Big[      
          + 4
          - 4H(0;y)
          + 4H(0;y)H(0;z)
          + 4H(0;y)H(1;z)
          - 6H(0;z)\nonumber \\
&&
          + H(0;z)H(1-z;y)
          + 4H(0,0;y)
          + 6H(0,0;z)
          - H(0,1;y)
          + 4H(0,1;z)\nonumber \\
&&
          + 2H(0,1-z;y)
          - 4H(1;z)
          + 2H(1;z)H(1-z;y)
          + 4H(1,0;z)
          + 4H(1,1;z)\nonumber \\
&&
          - 2H(1-z;y)
          + 2H(1-z,0;y)
          - H(1-z,1;y)
          + H(1-z,1-z;y) \Big] \; .
\end{eqnarray}
\begin{equation}
\boxbubblebNLO{p_{123}}{p_1}{p_2}{p_3} = \left(\frac{S_\e}{16
\pi^2}\right)^2\, \frac{\left( -\sabc \right)^{-2\e}}
{\sbc} \sum_{i=-1}^3 \frac{f_{5.2,i}
\left(\frac{\sac}{\sabc},\frac{\sbc}{\sabc}\right)}{\e^i}\; + 
{\cal O}(\e^2) ,
\end{equation}
with:
\begin{eqnarray}
f_{5.2,3}(y,z) &=& -1 \ , \\
f_{5.2,2}(y,z) &=& -2 + H(0;y)+ H(0;z) \ , \\
f_{5.2,1}(y,z) &=& 
          - 4
          + 2H(0;y)
          - H(0;y)H(0;z)
          + 2H(0;z)
          - 2H(0,0;y)
          - H(0,0;z)
          - H(1,0;y) \ ,  \\
f_{5.2,0}(y,z) &=& 
          - 8
          + 4H(0;y)
          - 2H(0;y)H(0;z)
          + H(0;y)H(1,0;z)
          + 4H(0;z)
          + H(0;z)H(1-z,0;y)\nonumber \\
&& 
          - 4H(0,0;y)
          + 2H(0,0;y)H(0;z)
          - 2H(0,0;z)
          + H(0,0;z)H(0;y)
          + 4H(0,0,0;y)\nonumber \\
&& 
          + H(0,0,0;z)
          + 2H(0,1,0;y)
          - 2H(1,0;y)
          + H(1,0;z)H(1-z;y)
          + 2H(1,0,0;y)\nonumber \\
&& 
          + H(1,1,0;z)
          + H(1-z,1,0;y) 
          + 5\zeta_3
      +\frac{\pi^2}{6} \left[
          + H(0;y)
          + H(1;z)
          + H(1-z;y) \right]  \ ,  \\
f_{5.2,-1}(y,z) &=& 
          - 16
          + 8H(0;y)
          - 4H(0;y)H(0;z)
          + 2H(0;y)H(1,0;z)
          - H(0;y)H(1,0,0;z)\nonumber \\
&& 
          - H(0;y)H(1,1,0;z)
          + 8H(0;z)
          + 2H(0;z)H(1-z,0;y)
          - 2H(0;z)H(1-z,0,0;y)\nonumber \\
&& 
          - H(0;z)H(1-z,1-z,0;y)
          - 8H(0,0;y)
          + 4H(0,0;y)H(0;z)
          - 2H(0,0;y)H(0,0;z)\nonumber \\
&& 
          - 2H(0,0;y)H(1,0;z)
          - 4H(0,0;z)
          + 2H(0,0;z)H(0;y)
          - H(0,0;z)H(1-z,0;y)\nonumber \\
&& 
          + 8H(0,0,0;y)
          - 4H(0,0,0;y)H(0;z)
          + 2H(0,0,0;z)
          - H(0,0,0;z)H(0;y)\nonumber \\
&& 
          - 8H(0,0,0,0;y)
          - H(0,0,0,0;z)
          - 4H(0,0,1,0;y)
          + 4H(0,1,0;y)\nonumber \\
&& 
          - H(0,1,0;z)H(0;y)
          - H(0,1,0;z)H(1-z;y)
          - 4H(0,1,0,0;y)
          - H(0,1,1,0;z)\nonumber \\
&& 
          - 2H(0,1-z;y)H(1,0;z)
          - 2H(0,1-z,0;y)H(0;z)
          - 2H(0,1-z,1,0;y)\nonumber \\
&& 
          - 4H(1,0;y)
          + 2H(1,0;z)H(1-z;y)
          - H(1,0;z)H(1-z,0;y)\nonumber \\
&& 
          - H(1,0;z)H(1-z,1-z;y)
          + 4H(1,0,0;y)
          - H(1,0,0;z)H(1-z;y)\nonumber \\
&& 
          - 4H(1,0,0,0;y)
          - H(1,0,1,0;z)
          + 2H(1,1,0;z)
          - H(1,1,0,0;z)\nonumber \\
&& 
          - 2H(1-z,0,1,0;y)
          + 2H(1-z,1,0;y)
          - 2H(1-z,1,0,0;y)\nonumber \\
&& 
          - H(1-z,1-z,1,0;y) 
          + \frac{37\pi^4}{360}
      + \zeta_3\left[
             10
          - 6H(0;y)
          - 5H(0;z)
          - H(1;z)
          - H(1-z;y)\right]\nonumber \\
&&
      + \frac{\pi^2}{6} \Big[
          + 2H(0;y)
          - H(0;y)H(0;z)
          - H(0;y)H(1;z)
          - H(0;z)H(1-z;y)\nonumber \\
&& 
          - 2H(0,0;y)
          - H(0,1;z)
          - 2H(0,1-z;y)
          + 2H(1;z)\nonumber \\
&& 
          - H(1,0;z)
          + 2H(1-z;y)
          - H(1-z,0;y)
          - H(1-z,1-z;y)
          \Big] \; .
\end{eqnarray}

The diagonal propagator insertion into the one-loop box with one off-shell 
leg gives rise to two different two-loop topologies, depending on 
whether the diagonal propagator is attached to the off-shell leg or not. 
One of these topologies contains one master integral, the other topology 
two master integrals. The choice of basis of master integrals 
for the latter
and the separation of the differential equations are discussed in 
Sections~\ref{sec:de} and~\ref{sec:example}. 

The master integrals for both topologies were computed in terms of 
generalised hypergeometric functions valid for arbitrary $\e$ in~\cite{gr}.
Again, an expansion in $\e$ 
of these functions is not straightforward, 
and we choose therefore to obtain the $\e$ expansions directly from 
the differential equations:
\begin{equation}
\boxxaNLO{p_{123}}{p_1}{p_2}{p_3} = \left(\frac{S_\e}{16
\pi^2}\right)^2\, \frac{\left( -\sabc \right)^{-2\e}}
{\sab} \sum_{i=0}^4 \frac{f_{5.3,i}
\left(\frac{\sac}{\sabc},\frac{\sbc}{\sabc}\right)}{\e^i}\; + 
{\cal O}(\e) ,
\end{equation}
with:
\begin{eqnarray}
f_{5.3,4}(y,z) &=& 0 \ , \\
f_{5.3,3}(y,z) &=& 0 \ , \\
f_{5.3,2}(y,z) &=& 
          - H(0;y)H(0;z)
          - H(1,0;y)
          - H(1,0;z) 
          - \frac{\pi^2}{6} \ , \\
f_{5.3,1}(y,z) &=& 
          + 2H(0;y)H(1,0;z)
          + 2H(0;z)H(1-z,0;y)
          + 2H(0,0;y)H(0;z)
          + 2H(0,0;z)H(0;y)\nonumber \\
&& 
          + 2H(0,1,0;y)
          + 2H(0,1,0;z)
          + 2H(1,0;z)H(1-z;y)
          + 2H(1,0,0;y)
          + 2H(1,0,0;z)\nonumber \\
&& 
          + 2H(1,1,0;z)
          + 2H(1-z,1,0;y)\nonumber \\
&& 
      + \frac{\pi^2}{6} \left[
          + 2H(0;y)
          + 2H(0;z)
          + 2H(1;z)
          + 2H(1-z;y) \right] \ ,  \\
f_{5.3,0}(y,z) &=& 
          - 4H(0;y)H(1,0,0;z)
          - 4H(0;y)H(1,1,0;z)
          - 4H(0;z)H(1-z,0,0;y)\nonumber \\
&& 
          - 4H(0;z)H(1-z,1-z,0;y)
          - 4H(0,0;y)H(0,0;z)
          - 4H(0,0;y)H(1,0;z)\nonumber \\
&& 
          - 4H(0,0;z)H(1-z,0;y)
          - 4H(0,0,0;y)H(0;z)
          - 4H(0,0,0;z)H(0;y)\nonumber \\
&& 
          - 4H(0,0,1,0;y)
          - 4H(0,0,1,0;z)
          - 4H(0,1,0;z)H(0;y)
          - 4H(0,1,0;z)H(1-z;y)\nonumber \\
&& 
          - 4H(0,1,0,0;y)
          - 4H(0,1,0,0;z)
          - 4H(0,1,1,0;z)
          - 4H(0,1-z;y)H(1,0;z)\nonumber \\
&& 
          - 4H(0,1-z,0;y)H(0;z)
          - 4H(0,1-z,1,0;y)
          - 4H(1,0;z)H(1-z,0;y)\nonumber \\
&& 
          - 4H(1,0;z)H(1-z,1-z;y)
          - 4H(1,0,0;z)H(1-z;y)
          - 4H(1,0,0,0;y)\nonumber \\
&& 
          - 4H(1,0,0,0;z)
          - 4H(1,0,1,0;z)
          - 4H(1,1,0;z)H(1-z;y)
          - 4H(1,1,0,0;z)\nonumber \\
&& 
          - 4H(1,1,1,0;z)
          - 4H(1-z,0,1,0;y)
          - 4H(1-z,1,0,0;y)
          - 4H(1-z,1-z,1,0;y)\nonumber \\
&& 
          - \frac{7\pi^4}{90} 
      + \frac{\pi^2}{6} \Big[
          - 4H(0;y)H(0;z)
          - 4H(0;y)H(1;z)
          - 4H(0;z)H(1-z;y)
          - 4H(0,0;y)\nonumber \\
&& 
          - 4H(0,0;z)
          - 4H(0,1;z)
          - 4H(0,1-z;y)
          - 4H(1;z)H(1-z;y)
          - 4H(1,0;z)\nonumber \\
&& 
          - 4H(1,1;z)
          - 4H(1-z,0;y)
          - 4H(1-z,1-z;y) \Big]  \ .
\end{eqnarray}
\begin{equation}
\boxxbNLO{p_{123}}{p_1}{p_2}{p_3} = \left(\frac{S_\e}{16
\pi^2}\right)^2\, \frac{\left( -\sabc \right)^{-2\e}}
{\sac+\sbc} \sum_{i=0}^4 \frac{f_{5.4,i}
\left(\frac{\sac}{\sabc},\frac{\sbc}{\sabc}\right)}{\e^i}\; + 
{\cal O}(\e) ,
\end{equation}
with:
\begin{eqnarray}
f_{5.4,4}(y,z) &=&  0  \ , \\
f_{5.4,3}(y,z) &=&  0  \ , \\
f_{5.4,2}(y,z) &=&  0  \ , \\
f_{5.4,1}(y,z) &=&  
          + H(0;y)H(1,0;z)
          + H(0;z)H(1-z,0;y)
          - H(0,1,0;y)
          - H(0,1,0;z)\nonumber \\
&& 
          + H(1,0;z)H(1-z;y)
          + H(1,1,0;z)
          + H(1-z,1,0;y)\nonumber \\
&& 
          + \frac{\pi^2}{6} \left[ + H(1;z)+ H(1-z;y) \right] \ ,  \\
f_{5.4,0}(y,z) &=&  
          - 2H(0;y)H(1,0,0;z)
          - H(0;y)H(1,1,0;z)
          - 2H(0;z)H(1-z,0,0;y)\nonumber \\
&& 
          - H(0;z)H(1-z,1-z,0;y)
          + 4H(0;z)H(z,1-z,0;y)
          - 2H(0,0;y)H(1,0;z)\nonumber \\
&& 
          - 2H(0,0;z)H(1-z,0;y)
          + 2H(0,0,1,0;y)
          - 2H(0,0,1,0;z)
          + 2H(0,1,0;z)H(0;y)\nonumber \\
&& 
          - 2H(0,1,0;z)H(1-z;y)
          - 4H(0,1,0;z)H(z;y)
          + 2H(0,1,0,0;y)
          + 2H(0,1,0,0;z)\nonumber \\
&& 
          - H(0,1,1,0;y)
          + H(0,1,1,0;z)
          - 2H(0,1-z;y)H(1,0;z)
          - 2H(0,1-z,0;y)H(0;z)\nonumber \\
&& 
          - 2H(0,1-z,1,0;y)
          - H(1,0;z)H(1-z,0;y)
          - H(1,0;z)H(1-z,1-z;y)\nonumber \\
&& 
          + 4H(1,0;z)H(z,0;y)
          + 4H(1,0;z)H(z,1-z;y)
          - 2H(1,0,0;z)H(1-z;y)\nonumber \\
&& 
          - 2H(1,0,1,0;z)
          + 4H(1,1,0;z)H(z;y)
          - 2H(1,1,0,0;z)
          - 2H(1-z,0,1,0;y)\nonumber \\
&& 
          - 2H(1-z,1,0,0;y)
          + H(1-z,1,1,0;y)
          - H(1-z,1-z,1,0;y)
          - 4H(z,0,1,0;y)\nonumber \\
&& 
          + 4H(z,1-z,1,0;y)
          + \zeta_3 \left[- 2H(1;z)- 2H(1-z;y)\right]\nonumber \\
&& 
      + \frac{\pi^2}{6} \Big[ 
          - H(0;y)H(1;z)
          - H(0;z)H(1-z;y)
          - H(0,1;y)
          + H(0,1;z)
          - 2H(0,1-z;y)\nonumber \\
&& 
          + 4H(1;z)H(z;y)
          - H(1,0;z)
          - H(1-z,0;y)
          + H(1-z,1;y)
          - H(1-z,1-z;y)\nonumber \\
&& 
          + 4H(z,1-z;y)  \Big] \ . 
\end{eqnarray}

\begin{equation}
\boxxbdotNLO{p_{123}}{p_1}{p_2}{p_3} = \left(\frac{S_\e}{16
\pi^2}\right)^2\, \frac{\left( -\sabc \right)^{-2\e}}
{\sac\sbc} \sum_{i=-1}^3 \frac{f_{5.5,i}
\left(\frac{\sac}{\sabc},\frac{\sbc}{\sabc}\right)}{\e^i}\; + 
{\cal O}(\e^2) ,
\end{equation}
with:
\begin{eqnarray}
f_{5.5,3}(y,z) &=&  1  \ , \\
f_{5.5,2}(y,z) &=&   - 2H(0;y)- 2H(0;z) \ , \\
f_{5.5,1}(y,z) &=&   
          + 4H(0;y)H(0;z)
          + 4H(0,0;y)
          + 4H(0,0;z)
          + 3H(1,0;y)
          + 3H(1,0;z) 
          + \frac{\pi^2}{3} \ , \\
f_{5.5,0}(y,z) &=&   
          - 6H(0;y)H(1,0;z)
          - 6H(0;z)H(1-z,0;y)
          - 8H(0,0;y)H(0;z)
          - 8H(0,0;z)H(0;y)\nonumber \\
&& 
          - 8H(0,0,0;y)
          - 8H(0,0,0;z)
          - 6H(0,1,0;y)
          - 6H(0,1,0;z)
          - 6H(1,0;z)H(1-z;y)\nonumber \\
&& 
          - 6H(1,0,0;y)
          - 6H(1,0,0;z)
          + 3H(1,1,0;y)
          - 3H(1,1,0;z)
          - 6H(1-z,1,0;y)\nonumber \\
&& 
          - 16 \zeta_3
    +\frac{\pi^2}{6} \left[ 
          - 4H(0;y)
          - 4H(0;z)
          + 3H(1;y)
          - 3H(1;z)
          - 6H(1-z;y) \right] \ ,  \\
f_{5.5,-1}(y,z) &=&   
          + 12H(0;y)H(1,0,0;z)
          + 6H(0;y)H(1,1,0;z)
          + 12H(0;z)H(1-z,0,0;y)\nonumber \\
&& 
          + 6H(0;z)H(1-z,1-z,0;y)
          + 16H(0,0;y)H(0,0;z)
          + 12H(0,0;y)H(1,0;z)\nonumber \\
&& 
          + 12H(0,0;z)H(1-z,0;y)
          + 16H(0,0,0;y)H(0;z)
          + 16H(0,0,0;z)H(0;y)\nonumber \\
&& 
          + 16H(0,0,0,0;y)
          + 16H(0,0,0,0;z)
          + 12H(0,0,1,0;y)
          + 12H(0,0,1,0;z)\nonumber \\
&& 
          + 12H(0,1,0;z)H(0;y)
          + 12H(0,1,0;z)H(1-z;y)
          + 12H(0,1,0,0;y)\nonumber \\
&& 
          + 12H(0,1,0,0;z)
          - 6H(0,1,1,0;y)
          + 6H(0,1,1,0;z)
          + 12H(0,1-z;y)H(1,0;z)\nonumber \\
&& 
          + 12H(0,1-z,0;y)H(0;z)
          + 12H(0,1-z,1,0;y)
          + 6H(1,0;z)H(1-z,0;y)\nonumber \\
&& 
          + 6H(1,0;z)H(1-z,1-z;y)
          + 12H(1,0,0;z)H(1-z;y)
          + 12H(1,0,0,0;y)\nonumber \\
&& 
          + 12H(1,0,0,0;z)
          + 12H(1,0,1,0;z)
          - 6H(1,1,0,0;y)
          + 6H(1,1,0,0;z)\nonumber \\
&& 
          + 3H(1,1,1,0;y)
          + 3H(1,1,1,0;z)
          + 12H(1-z,0,1,0;y)
          + 12H(1-z,1,0,0;y)\nonumber \\
&& 
          - 6H(1-z,1,1,0;y)
          + 6H(1-z,1-z,1,0;y) 
          - \frac{2\pi^4}{5}\nonumber \\
&& 
   + \zeta_3\left[
          + 32H(0;y)
          + 32H(0;z)
          + 3H(1;y)
          + 15H(1;z)
          + 12H(1-z;y) \right]\nonumber \\
&& 
    +\frac{\pi^2}{6} \Big[    
          + 8H(0;y)H(0;z)
          + 6H(0;y)H(1;z)
          + 6H(0;z)H(1-z;y)
          + 8H(0,0;y)\nonumber \\
&& 
          + 8H(0,0;z)
          - 6H(0,1;y)
          + 6H(0,1;z)
          + 12H(0,1-z;y)
          + 6H(1,0;z)
          + 3H(1,1;y)\nonumber \\
&& 
          + 3H(1,1;z)
          + 6H(1-z,0;y)
          - 6H(1-z,1;y)
          + 6H(1-z,1-z;y) \Big] \ . 
\end{eqnarray}

The only six propagator planar master integral was up to now not known in the 
literature:
\begin{equation}
\boxxbpNLO{p_{123}}{p_1}{p_2}{p_3} = \left(\frac{S_\e}{16
\pi^2}\right)^2\, \frac{\left( -\sabc \right)^{-2\e}}
{(\sab+\sbc)\sbc} \sum_{i=0}^4 \frac{f_{6.1,i}
\left(\frac{\sac}{\sabc},\frac{\sbc}{\sabc}\right)}{\e^i}\; + 
{\cal O}(\e) ,
\end{equation}
with:
\begin{eqnarray}
f_{6.1,4}(y,z) &=&  0  \ , \\
f_{6.1,3}(y,z) &=&  -\frac{1}{2} H(0;y) \ ,  \\
f_{6.1,2}(y,z) &=&  
          + H(0,0;y)
          - H(1,0;y) 
          - \frac{\pi^2}{6} \ , \\
f_{6.1,1}(y,z) &=& 
          + H(0;y)H(1,0;z)
          + H(0;z)H(1-z,0;y)
          + H(0,0;z)H(0;y)
          - 2H(0,0,0;y)\nonumber \\
&& 
          - H(0,1,0;y)
          + H(0,1,0;z)
          + H(1,0;z)H(1-z;y)
          + 2H(1,0,0;y)
          - 2H(1,1,0;y)\nonumber \\
&& 
          + H(1,1,0;z)
          + H(1-z,1,0;y)
          - 3\zeta_3\nonumber \\
&& 
  + \frac{\pi^2}{6} \left[
          + H(0;z)
          - 2H(1;y)
          + H(1;z)
          + H(1-z;y) \right] \ , \\
f_{6.1,0}(y,z) &=& 
          - 3H(0;y)H(1,0,0;z)
          - H(0;y)H(1,1,0;z)
          + 2H(0;z)H(1,1-z,0;y)\nonumber \\
&& 
          - 2H(0;z)H(1-z,0,0;y)
          - H(0;z)H(1-z,1-z,0;y)
          - 2H(0,0;y)H(0,0;z)\nonumber \\
&& 
          - 2H(0,0;y)H(1,0;z)
          + 2H(0,0;z)H(1,0;y)
          - 3H(0,0;z)H(1-z,0;y)\nonumber \\
&& 
          - 3H(0,0,0;z)H(0;y)
          + 4H(0,0,0,0;y)
          + 2H(0,0,1,0;y)
          - 3H(0,0,1,0;z)\nonumber \\
&& 
          - H(0,1,0;z)H(0;y)
          + 2H(0,1,0;z)H(1;y)
          - 3H(0,1,0;z)H(1-z;y)
          + 2H(0,1,0,0;y)\nonumber \\
&& 
          - 3H(0,1,0,0;z)
          + H(0,1,1,0;y)
          - 2H(0,1-z;y)H(1,0;z)
          - 2H(0,1-z,0;y)H(0;z)\nonumber \\
&& 
          - 2H(0,1-z,1,0;y)
          + 2H(1,0;y)H(1,0;z)
          + 2H(1,0;z)H(1,1-z;y)\nonumber \\
&& 
          - H(1,0;z)H(1-z,0;y)
          - H(1,0;z)H(1-z,1-z;y)
          - 3H(1,0,0;z)H(1-z;y)\nonumber \\
&& 
          - 4H(1,0,0,0;y)
          - 2H(1,0,1,0;y)
          - 3H(1,0,1,0;z)
          + 2H(1,1,0;z)H(1;y)\nonumber \\
&& 
          + 4H(1,1,0,0;y)
          - 3H(1,1,0,0;z)
          - 4H(1,1,1,0;y)
          + 2H(1,1-z,1,0;y)\nonumber \\
&& 
          - 2H(1-z,0,1,0;y)
          - 2H(1-z,1,0,0;y)
          + 2H(1-z,1,1,0;y)
          - H(1-z,1-z,1,0;y)  \nonumber \\
&&           
          + \frac{\pi^4}{72}
  + \zeta_3\left[ 
          - 2H(0;y)
          - 3H(0;z)
          - 6H(1;y)
          - 3H(1;z)
          - 3H(1-z;y) \right]\nonumber \\
&& 
    +\frac{\pi^2}{6} \Big[    
          - H(0;y)H(0;z)
          - H(0;y)H(1;z)
          + 2H(0;z)H(1;y)
          - H(0;z)H(1-z;y)\nonumber \\
&& 
          - H(0,0;z)
          + H(0,1;y)
          - 2H(0,1-z;y)
          + 2H(1;y)H(1;z)
          - H(1,0;z)
          - 4H(1,1;y)\nonumber \\
&& 
          + 2H(1,1-z;y)
          - H(1-z,0;y)
          + 2H(1-z,1;y)
          - H(1-z,1-z;y) \Big] \ .
\end{eqnarray}

The seven propagator planar double box topology contains two master integrals. 
We choose the basis first suggested in~\cite{onshell6} for the 
double box integrals with all legs on-shell, which yields particularly 
compact results also in our case with one leg off-shell:
\begin{equation}
\doubleboxNLO{p_{123}}{p_1}{p_2}{p_3} = \left(\frac{S_\e}{16
\pi^2}\right)^2\, \frac{\left( -\sabc \right)^{-2\e}}
{\sac\sbc^2} \sum_{i=0}^4 \frac{f_{7.1,i}
\left(\frac{\sac}{\sabc},\frac{\sbc}{\sabc}\right)}{\e^i}\; + 
{\cal O}(\e) ,
\label{eq:master73}
\end{equation}
with:
\begin{eqnarray}
f_{7.1,4}(y,z) &=& -1 \ , \\
f_{7.1,3}(y,z) &=& 2\left[H(0;y)+H(0;z)\right] \ , \\
f_{7.1,2}(y,z) &=&           - 4H(0;y)H(0;z)
          - 4H(0,0;y)
          - 4H(0,0;z)
          - H(1,0;y)
          - 3H(1,0;z) -\frac{7\pi^2}{12} \ , \\
f_{7.1,1}(y,z) & = & 
          + 6H(0;y)H(1,0;z)
          + 2H(0;z)H(1-z,0;y)
          + 8H(0,0;y)H(0;z)\nonumber \\
&&
          + 8H(0,0;z)H(0;y)
          + 8H(0,0,0;y)
          + 8H(0,0,0;z)
          + 5H(0,1,0;y)\nonumber\\
&&
          + 6H(0,1,0;z)
          + 2H(1,0;z)H(1-z;y)
          + 2H(1,0,0;y)
          + 9H(1,0,0;z)\nonumber \\
&&
          - 2H(1,1,0;y)
          + 6H(1,1,0;z)
          + 2H(1-z,1,0;y)
          - \frac{\zeta_3}{2}\nonumber \\
&&
      +\frac{\pi^2}{6} \left[
          + 7H(0;y)
          + 7H(0;z)
          - 2H(1;y)
          + 6H(1;z)
          + 2H(1-z;y) \right] \ , \\
f_{7.1,0}(y,z) & = & 
          - 18H(0;y)H(1,0,0;z)
          - 12H(0;y)H(1,1,0;z)
          + 2H(0;z)H(1,1-z,0;y)\nonumber \\
&&
          - 4H(0;z)H(1-z,0,0;y)
          - 2H(0;z)H(1-z,1-z,0;y)
          - 16H(0,0;y)H(0,0;z)\nonumber \\
&&
          - 12H(0,0;y)H(1,0;z)
          + 2H(0,0;z)H(1,0;y)
          - 6H(0,0;z)H(1-z,0;y)\nonumber \\
&&
          - 16H(0,0,0;y)H(0;z)
          - 16H(0,0,0;z)H(0;y)
          - 16H(0,0,0,0;y)\nonumber \\
&&
          - 16H(0,0,0,0;z)
          - 16H(0,0,1,0;y)
          - 12H(0,0,1,0;z)
          - 12H(0,1,0;z)H(0;y)\nonumber \\
&&
          + 2H(0,1,0;z)H(1;y)
          - 6H(0,1,0;z)H(1-z;y)
          - 10H(0,1,0,0;y)\nonumber \\
&&
          - 18H(0,1,0,0;z)
          + H(0,1,1,0;y)
          - 12H(0,1,1,0;z)
          - 10H(0,1-z;y)H(1,0;z)\nonumber \\
&&
          - 10H(0,1-z,0;y)H(0;z)
          - 10H(0,1-z,1,0;y)
          + 2H(1,0;y)H(1,0;z)\nonumber \\
&&
          + 2H(1,0;z)H(1,1-z;y)
          - 4H(1,0;z)H(1-z,0;y)
          - 2H(1,0;z)H(1-z,1-z;y)\nonumber \\
&&
          - 6H(1,0,0;z)H(1-z;y)
          - 4H(1,0,0,0;y)
          - 21H(1,0,0,0;z)
          - 2H(1,0,1,0;y)\nonumber \\
&&
          - 15H(1,0,1,0;z)
          + 2H(1,1,0;z)H(1;y)
          - 4H(1,1,0;z)H(1-z;y)\nonumber \\
&&
          + 4H(1,1,0,0;y)
          - 21H(1,1,0,0;z)
          - 4H(1,1,1,0;y)
          - 12H(1,1,1,0;z)\nonumber \\
&&
          + 2H(1,1-z,1,0;y)
          - 2H(1-z,0,1,0;y)
          - 4H(1-z,1,0,0;y)\nonumber \\
&&
          + 4H(1-z,1,1,0;y)
          - 2H(1-z,1-z,1,0;y)
          - \frac{\pi^4}{4}\nonumber \\
&&
      + \zeta_3\left[
          + H(0;y)
          - 2H(1-z;y)
          - 6H(1;y)
          + 3H(1;z)
          + H(0;z)\right]\nonumber \\
&&
      + \frac{\pi^2}{6} \Big[
          - 14H(0;y)H(0;z)
          - 12H(0;y)H(1;z)
          + 2H(0;z)H(1;y)
          - 2H(0;z)H(1-z;y)\nonumber \\
&&
          - 14H(0,0;y)
          - 14H(0,0;z)
          + H(0,1;y)
          - 12H(0,1;z)
          - 10H(0,1-z;y)\nonumber \\
&&
          + 2H(1;y)H(1;z)
          - 4H(1;z)H(1-z;y)
          - 15H(1,0;z)
          - 4H(1,1;y)
          - 12H(1,1;z)\nonumber \\
&&
          + 2H(1,1-z;y)
          - 4H(1-z,0;y)
          + 4H(1-z,1;y)
          - 2H(1-z,1-z;y) \Big] \; .
\end{eqnarray}

\begin{equation}
\doubleboxNLOtwo{p_{123}}{p_1}{p_2}{p_3} = \left(\frac{S_\e}{16
\pi^2}\right)^2\, \frac{\left( -\sabc \right)^{-2\e}}
{(\sab+\sac)\sbc} \sum_{i=0}^4 \frac{f_{7.2,i}
\left(\frac{\sac}{\sabc},\frac{\sbc}{\sabc}\right)}{\e^i} + 
{\cal O}(\e)\; ,
\end{equation}
with:
\begin{eqnarray}
f_{7.2,4}(y,z) &=& 0 \ , \\
f_{7.2,3}(y,z) &=& -\frac{3}{2} H(0;z) \ ,  \\
f_{7.2,2}(y,z) &=&           
          + 2H(0;y)H(0;z)
          + \frac{9}{2}H(0,0;z)
          + \frac{5}{2}H(1,0;z)
          + \frac{5\pi^2}{12} \ ,  \\
f_{7.2,1}(y,z) &=& 
          - 4H(0;y)H(1,0;z)
          - 4H(0;z)H(1-z,0;y)
          - 4H(0,0;y)H(0;z)
          - 6H(0,0;z)H(0;y)\nonumber \\
&&
          - \frac{21}{2}H(0,0,0;z)
          - 4H(0,1,0;y)
          - \frac{11}{2}H(0,1,0;z)
          - 4H(1,0;z)H(1-z;y)\nonumber \\
&&
          - \frac{15}{2}H(1,0,0;z)
          - \frac{7}{2}H(1,1,0;z)
          - 4H(1-z,1,0;y) \nonumber \\
&& + \frac{\pi^2}{6} \left[           
          - 4H(0;y)
          - \frac{15}{2}H(0;z)
          - \frac{7}{2}H(1;z)
          - 4H(1-z;y) \right] \ , \\
f_{7.2,0}(y,z) &=& 
          + 12H(0;y)H(1,0,0;z)
          + 8H(0;y)H(1,1,0;z)
          + 8H(0;z)H(1-z,0,0;y)\nonumber \\
&&
          + 10H(0;z)H(1-z,1-z,0;y)
          + 12H(0,0;y)H(0,0;z)
          + 8H(0,0;y)H(1,0;z)\nonumber \\
&&
          + 12H(0,0;z)H(1-z,0;y)
          + 8H(0,0,0;y)H(0;z)
          + 14H(0,0,0;z)H(0;y)\nonumber \\
&&
          + \frac{45}{2}H(0,0,0,0;z)
          + 10H(0,0,1,0;y)
          + \frac{23}{2}H(0,0,1,0;z)
          + 12H(0,1,0;z)H(0;y)\nonumber \\
&&
          + 12H(0,1,0;z)H(1-z;y)
          + 8H(0,1,0,0;y)
          + \frac{27}{2}H(0,1,0,0;z)
          - 2H(0,1,1,0;y)\nonumber \\
&&
          + \frac{21}{2}H(0,1,1,0;z)
          + 10H(0,1-z;y)H(1,0;z)
          + 10H(0,1-z,0;y)H(0;z)\nonumber \\
&&
          + 10H(0,1-z,1,0;y)
          + 8H(1,0;z)H(1-z,0;y)
          + 10H(1,0;z)H(1-z,1-z;y)\nonumber \\
&&
          + 12H(1,0,0;z)H(1-z;y)
          + \frac{35}{2}H(1,0,0,0;z)
          + \frac{25}{2}H(1,0,1,0;z)\nonumber \\
&&
          + 8H(1,1,0;z)H(1-z;y)
          + \frac{21}{2}H(1,1,0,0;z)
          + \frac{17}{2}H(1,1,1,0;z)
          + 10H(1-z,0,1,0;y)\nonumber \\
&&
          + 8H(1-z,1,0,0;y)
          - 2H(1-z,1,1,0;y)
          + 10H(1-z,1-z,1,0;y)
          + \frac{37 \pi^4}{240} \nonumber \\
&&
+ \zeta_3 \left[    
          + 4H(0;y)
          + 6H(1;z)
          + 4H(1-z;y) \right] \nonumber \\
&&
+ \frac{\pi^2}{6} \bigg[       
          + 10H(0;y)H(0;z)
          + 8H(0;y)H(1;z)
          + 10H(0;z)H(1-z;y)
          + 8H(0,0;y)\nonumber \\
&&
          + \frac{35}{2}H(0,0;z)
          - 2H(0,1;y)
          + \frac{21}{2}H(0,1;z)
          + 10H(0,1-z;y)
          + 8H(1;z)H(1-z;y)\nonumber \\
&&
          + \frac{21}{2}H(1,0;z)
          + \frac{17}{2}H(1,1;z)
          + 8H(1-z,0;y)
          - 2H(1-z,1;y)
          + 10H(1-z,1-z;y)
            \bigg] \ . \nonumber \\
\end{eqnarray}

The first of the three-scale double box 
integrals (\ref{eq:master73}), was recently computed by Smirnov~\cite{smirnew}
using a Mellin-Barnes representation. We confirm this result. Since 
a comparison of the result of~\cite{smirnew} with our result 
is not straightforward, we  outline below the procedure 
we have employed.

In~\cite{smirnew} analytic forms containing Nielsen's polylogarithms 
are given for all $\e$-divergent terms of (\ref{eq:master73}); 
these terms can be compared directly. For the finite part of
(\ref{eq:master73}), Smirnov provides in~\cite{smirnew}
a result involving a
one-dimensional integral plus other terms containing only 
Nielsen's polylogarithms of non-simple 
arguments. The integrand of the one-dimensional integral 
involves products of logarithms with dilogarithms. 

To compare with our result, we rewrote the finite part quoted 
in~\cite{smirnew} in terms of 2dHPL. The procedure employed was as
follows. In the limit $y\to 0$, we determined the $z$-dependent coefficients 
of $\ln^i y$ ($i=1\ldots 4$) as well as the finite $z$-dependent term. 
The remaining terms, which are 
regular in $y \to 0$ were determined from the $y$-derivative of 
the result. Taking this 
derivative, one obtains a combination of rational fractions 
$1/y,1/(1-y),1/(1-y-z)$ times 
Nielsen's polylogarithms of level 3. These polylogarithms are
differentiated with respect to $y$ again, resulting in 
 $(1/y,1/(1-y),1/(1-y-z))$ times 
Nielsen's polylogarithms of level 2, which are translated into 
2dHPL of weight 2. The two $y$ derivatives are undone by integration
(keeping correct account of boundary terms by verifying the limit $y\to
0$ at each stage). As a result, we obtain 2dHPL of weight 4, as well as 
products of 2dHPL with ordinary HPL of $z$. After converting the
overall normalisation factor, we find  agreement 
of~\cite{smirnew} with our 
result (\ref{eq:master73}).

\section{Conclusion}
\label{sec:conc}
\setcounter{equation}{0}

Two-loop four-point functions with one off-shell leg are an important 
ingredient to the calculation of next-to-next-to-leading order 
corrections to three jet production and related observables 
in electron-positron annihilation. By exploiting 
integration-by-parts~\cite{hv,chet1,chet2} and Lorentz invariance~\cite{gr} 
identities, one can express the loop integrals appearing in these 
functions as a linear combination of a small set of master 
integrals. These master integrals are scalar 
functions of the external invariants, their determination was 
up to now a major obstacle to further progress in
next-to-next-to-leading order calculations. 

In the present paper, we compute all master integrals appearing in the 
reduction of planar two-loop four-point functions with 
massless internal propagators and one off-shell
leg. The method employed here relies on the fact that all master
integrals fulfil inhomogeneous differential equations in their external
invariants. For the master integrals under 
consideration, these differential equations were derived in~\cite{gr}. 
We determined the master integrals by solving the corresponding
differential equations, starting from a product ansatz involving 
a rational function of the external invariants and 
a sum of newly introduced
two-dimensional harmonic polylogarithms. The 
two-dimensional harmonic polylogarithms are an extension of the 
harmonic polylogarithms of~\cite{hpl}, built {\it \a`{a} la carte}
 to suit the needs of 
our calculation. We describe the underlying formalism in the Appendix
and tabulate the relation of  two-dimensional harmonic polylogarithms to 
the better known generalised polylogarithms of Nielsen~\cite{nielsen}. 
For the divergent 
parts of the master integrals considered here, we find analytic
expressions in terms of generalised polylogarithms of weight up to 3, 
while the finite part involves also 2dHPL of weight 4 which 
can be expressed as a one-dimensional 
integral over a combination of generalised polylogarithms of weight 3 
with suitable arguments. 

The application of our results to different kinematic situations, as 
encountered in the next-to-next-to-leading order corrections to 
vector boson plus jet production at hadron colliders or two plus one
jet production in deep inelastic scattering requires the analytic continuation 
of the two-dimensional harmonic polylogarithms outside the 
region $0\leq y\leq 1$, $0 \leq z \leq 1-y$, where they are real. 
An algorithm to derive these analytic continuation formulae is 
outlined in the Appendix.

A yet outstanding task is the computation of the master integrals
appearing in the reduction of amplitudes with crossed topologies. 
The derivation of these from the corresponding differential equations 
should follow the procedure outlined in Section~\ref{sec:de}. The ansatz 
does however contain more than one rational function ${\cal R}_i$, as 
already observed in the on-shell case~\cite{onshell3,onshell4}.
Moreover, the determination of the boundary conditions is more 
involved since the non-planar topologies have in general three branch
points: in $y=0$, $z=0$ and $y=1-z$. 
This work will be reported in a separate publication~\cite{nplan}. 

\section*{Acknowledgement} 
We are grateful to Jos Vermaseren for his assistance in the use of 
the algebraic program FORM~\cite{form} which was intensively used in 
all the steps of the calculation. 

\begin{appendix}
\renewcommand{\theequation}{\mbox{\Alph{section}.\arabic{equation}}}

\section{Harmonic Polylogarithms}
\setcounter{equation}{0}

Harmonic polylogarithms (HPL) were introduced in~\cite{hpl} as an
extension of the generalised polylogarithms of Nielsen~\cite{nielsen}. 
They are 
constructed in such a way that they form a closed, linearly 
independent set under a certain class of integrations. We observe that the 
class of allowed integrations on this set can be extended {\it \a`{a} la
carte} by enlarging the definition of harmonic polylogarithms
in order to suit the needs of a particular calculation. We will make 
use of this feature by generalising the one-dimensional HPL
of~\cite{hpl} to two-dimensional harmonic polylogarithms (2dHPL),
which appear in the solution of the differential equations for 
the three-scale master integrals discussed in this paper. 

In the following subsection, we will briefly review the properties of the 
ordinary HPL, and tabulate expressions for HPL up to weight 4 in terms 
of  generalised polylogarithms. In another subsection, 
we will then extend this 
framework to 2dHPL, which can in general be
expressed up to weight 3 in terms of generalised polylogarithms of 
suitable non-simple arguments. We tabulate
all 2dHPL up to weight 3. The 2dHPL of weight 4 are 
expressed as one-dimensional integrals over  
combinations
of generalised polylogarithms and therefore 
can be evaluated numerically in a 
straightforward manner.

\subsection{One-dimensional Harmonic Polylogarithms}

The one-dimensional HPL $H(\vec{m}_w;x)$ is described by a $w$-dimensional 
vector $\vec{m}_w$ of parameters and by its argument $x$. $w$ is called 
the weight of $H$. We briefly recall the 
HPL formalism:
\begin{enumerate}
\item
Definition of the  HPL at $w=1$:
\begin{eqnarray}
H(1;x) & \equiv & -\ln (1-x)\; ,\nonumber \\
H(0;x) & \equiv & \ln x \; ,\nonumber \\
H(-1;x) & \equiv & \ln (1+x)
\label{eq:levelone}
\end{eqnarray}
and the rational fractions in $x$
\begin{eqnarray}
f(1;x) & \equiv & \frac{1}{1-x} \;, \nonumber \\
f(0;x) & \equiv & \frac{1}{x} \;, \nonumber \\
f(-1;x) & \equiv & \frac{1}{1+x} \;,
\end{eqnarray}
such that 
\begin{equation}
\frac{\partial}{\partial x} H(a;x) = f(a;x)\qquad \mbox{with}\quad
a=+1,0,-1\;.
\end{equation}
\item For $w>1$:
\begin{eqnarray}
H(0,\ldots,0;x) & \equiv & \frac{1}{w!} \ln^w x\; ,\\
H(a,\vec{b};x) & \equiv & \int_0^x \d x^{\prime} f(a;x^{\prime}) H(\vec{b};x^{\prime})\; , 
\label{eq:inth}
\end{eqnarray}
which results in 
\begin{equation}
\frac{\partial}{\partial x} H(a,\vec{b};x) = f(a;x) H(\vec{b};x)\;.
\label{eq:derivh}
\end{equation}
This last relation is the main tool for verifying identities among 
different HPL. Such identities can be verified by first checking a 
special point (typically $x=0$)
and subsequently checking the derivatives. If agreement in 
the derivatives is not obvious, this procedure can be repeated until one 
arrives at relations involving only HPL with $w=1$.
\item
The HPL fulfil an algebra (see Section 3 of~\cite{hpl}), such that 
a product of two HPL (with weights $w_1$ and $w_2$) 
of the same argument $x$ is a combination of HPL of argument 
$x$ with weight $w=w_1+w_2$, 
\begin{equation}
H(\vec{a};x) H(\vec{b};x) = \sum_{\vec{c} = \vec{a} \uplus \vec{b}}
H(\vec{c};x),
\label{eq:alg}
\end{equation}
where $\vec{a} \uplus \vec{b}$ represents all mergers of $\vec{a}$ and 
$\vec{b}$ in which the relative orders of the elements of $\vec{a}$ and 
$\vec{b}$  are preserved. 
For example at $w_1=1,w_2=3$, this relation 
reads:
\begin{eqnarray}
H(m_1;x) H(m_2,m_3,m_4;x) &=& 
+H(m_1,m_2,m_3,m_4;x) + H(m_2,m_1,m_3,m_4;x) \nonumber \\
       &&        + H(m_2,m_3,m_1,m_4;x) + H(m_2,m_3,m_4,m_1;x) \; .
\label{eq:exal}
\end{eqnarray}
\item The HPL fulfil the integration-by-parts identities (see also Section 
3 of~\cite{hpl})
\begin{eqnarray}
H(m_1,\ldots,m_q;x) &=&  H(m_1;x)H(m_2,\ldots,m_q;x)
                        -H(m_2,m_1;x)H(m_3,\ldots,m_q;x) \nonumber \\
&& + \ldots + (-1)^{q+1} H(m_q,\ldots,m_1;x)\;.
\label{eq:ibp}
\end{eqnarray} 
\item
The HPL are linearly independent.
\end{enumerate}

The product identities (\ref{eq:alg}) and the integration-by-parts 
identities (\ref{eq:ibp}) involve different polylogarithms 
of the same weight $w$, as well as products of 
harmonic polylogarithms of lower weight. They can thus be used to 
express all polylogarithms of weight $w$ in terms of a so-called 'minimal 
set' of weight $w$ plus products of HPL of lower weight.  

It can be seen from (\ref{eq:inth}), that the HPL of parameters $(+1,0,-1)$ 
form a closed set under the class of integrations
\begin{equation}
\int_0^x \d x^{\prime} \left(\frac{1}{x^{\prime}}, 
\frac{1}{1-x^{\prime}},\frac{1}{1+x^{\prime}}\right) 
H(\vec{b};x^{\prime})\; .
\label{eq:intclass}
\end{equation}
In the context of the present calculation, it turns out 
that integrals 
involving denominators $1/(1+x)$ do never occur, the HPL of parameters 
$(0,1)$ are therefore sufficient. 

The HPL of parameters $(0,1)$ can all be expressed in 
terms of logarithms and 
Nielsen's polylogarithms. Up to level $m=4$, they read:
\begin{eqnarray}
  H(0;x) &=& \ln x\ , \nonumber \\
  H(1;x) &=& -\ln (1-x)\ , \nonumber \\
  H(0,0;x) &=& \frac{1}{2!} \ln^2x \ , \nonumber\\
  H(0,1;x) &=& \Li_2(x) \ , \nonumber\\
  H(1,0;x) &=& - \ln{x} \ln(1-x) - \Li_2(x) \ , \nonumber\\
  H(1,1;x) &=& \frac{1}{2!} \ln^2(1-x) \ , \nonumber \\
  H(0,0,0;x) &=& \frac{1}{3!} \ln^3x \ , \nonumber\\
  H(0,0,1;x) &=& \Li_3(x) \ , \nonumber\\
  H(0,1,0;x) &=& -2\Li_3(x) + \ln x \Li_2(x) \ , \nonumber\\
  H(0,1,1;x) &=& \S_{1,2}(x) \ , \nonumber\\
  H(1,0,0;x) &=& -\frac{1}{2} \ln (1-x)\ln^2x - \ln x \Li_2(x) + \Li_3(x)
                   \ , \nonumber\\
  H(1,0,1;x) &=& -2\S_{1,2}(x) - \ln(1-x)\Li_2(x)  \ , \nonumber\\
  H(1,1,0;x) &=& S_{1,2}(x) + \ln(1-x)\Li_2(x) + \frac{1}{2}\ln x\ln^2(1-x)
                          \ , \nonumber\\
  H(1,1,1;x) &=& -\frac{1}{3!} \ln^3(1-x) \ , \nonumber\\
  H(0,0,0,0;x) &=& \frac{1}{4!} \ln^4x \ , \nonumber\\
  H(0,0,0,1;x) &=& \Li_4(x) \ , \nonumber\\
  H(0,0,1,0;x) &=& \ln x \Li_3(x) - 3\Li_4(x)  \ , \nonumber\\
  H(0,0,1,1;x) &=& \S_{2,2}(x) \ , \nonumber\\
  H(0,1,0,0;x) &=& \frac{1}{2} \ln^2x \Li_2(x) - 2\ln x \Li_3(x) + 3\Li_4(x)
                   \ , \nonumber\\
  H(0,1,0,1;x) &=& -2\S_{2,2}(x) + \frac{1}{2} [\Li_2(x)]^2  \ ,\nonumber\\
  H(0,1,1,0;x) &=& \ln x S_{1,2}(x) -\frac{1}{2} [\Li_2(x)]^2  \ ,\nonumber\\
  H(0,1,1,1;x) &=& \S_{1,3}(x) \ , \nonumber\\
  H(1,0,0,0;x) &=& -\frac{1}{6}\ln^3x\ln(1-x) - \frac{1}{2}\ln^2x\Li_2(x) 
                   + \ln(x)\Li_3(x) - \Li_4(x)  \ , \nonumber\\
  H(1,0,0,1;x) &=& -\frac{1}{2} [\Li_2(x)]^2  - \ln(1-x)\Li_3(x)
                   \ , \nonumber\\
  H(1,0,1,0;x) &=& 2\ln(1-x)\Li_3(x) - \ln x \ln(1-x) \Li_2(x) 
                   -2\ln x\S_{1,2}(x) +  \frac{1}{2} [\Li_2(x)]^2 
                   + 2\S_{2,2}(x)  \ , \nonumber\\
  H(1,0,1,1;x) &=& -\ln (1-x) \S_{1,2}(x) -3 \S_{1,3}(x) \ , \nonumber\\
  H(1,1,0,0;x) &=& \frac{1}{4} \ln^2x\ln^2(1-x) - \ln(1-x)\Li_3(x) 
                   + \ln x\ln(1-x) \Li_2(x) + \ln x\S_{1,2}(x)-\S_{2,2}(x)
                   \ , \nonumber\\
  H(1,1,0,1;x) &=& \frac{1}{2} \ln^2(1-x) \Li_2(x) + 2\ln(1-x)\S_{1,2}(x)  
                   + 3 \S_{1,3}(x) \ , \nonumber\\
  H(1,1,1,0;x) &=& -\frac{1}{6} \ln x \ln^3(1-x) 
                   - \frac{1}{2}\ln^2(1-x)\Li_2(x) 
                   - \ln (1-x) \S_{1,2}(x) -  S_{1,3}(x) \ , \nonumber\\
  H(1,1,1,1;x) &=& \frac{1}{4!} \ln^4(1-x) \ .
\end{eqnarray}

\subsection{Two-dimensional Harmonic Polylogarithms}
\label{ap:two}

The generalisation from one-dimensional to two-dimensional 
HPL starts from (\ref{eq:intclass}), which defines the class of 
integrations under which the HPL form a closed set. By inspection of 
the various 
 inhomogeneous terms of the $y$ differential equations for the 
three-scale master integrals discussed in this paper, we find that,
besides the denominators $1/y$ and $1/(1-y)$ also $1/(1-y-z)$ and 
$1/(y+z)$ appear. It is therefore appropriate to introduce an 
extension of the HPL, which forms a closed set under the class 
of integrations 
\begin{equation}
\int_0^y \d y^{\prime} \left(\frac{1}{y^{\prime}}, 
  \frac{1}{1-y^{\prime}},\frac{1}{1-y^{\prime}-z},\frac{1}{y^{\prime}+z}
\right) H(\vec{b};y^{\prime})\; .
\label{eq:twoddef}
\end{equation}
Such an extension of the HPL formalism 
can be made by extending the 
set of fractions by
\begin{eqnarray}
f(1-z;y) & \equiv & \frac{1}{1-y-z} \;, \nonumber \\
f(z;y) & \equiv & \frac{1}{y+z} \;, 
\end{eqnarray}
and correspondingly the set of HPL at weight 1 by
\begin{eqnarray}
H(1-z;y) &=& - \ln\left(1-\frac{y}{1-z}\right) \ , \nonumber \\
H(z;y) &=&  \ln\left(\frac{y+z}{z}\right) .
\end{eqnarray}
Allowing $(z,1-z)$ as components of the 
vector $\vec{m}_w$ of parameters, (\ref{eq:inth}) does then define the 
extended set of HPL, which we call two-dimensional 
harmonic polylogarithms (2dHPL).  They form a closed set  
under integrations of the form (\ref{eq:twoddef}), as required. 
This closure was achieved by construction, showing that the
HPL formalism of~\cite{hpl} can be extended to suit the needs 
of a particular calculation. We do not consider $-1$ among  the
components of $\vec{m}_w$, since they are not needed in the 
present context. The 2dHPL fulfil the same algebra as the HPL (\ref{eq:alg}), 
and they are 
linearly independent. The feature of linear independence distinguishes
the 2dHPL from Nielsen's polylogarithms of 
non-trivial  two-variable arguments
(in the literature, see for
example~\cite{lewin},
there are large numbers of relations between Nielsen's polylogarithms 
with arguments which are
rational functions of two variables). 
Let us observe here that the use of linearly independent functions 
protects against the danger of writing ``hidden zeroes" (complicated 
expressions whose actual value is identically zero), making it 
trivial at the same time to carry out those simplifications whose occurrence 
is among the main features of any analytical calculation. 
Finally, the integration-by-parts relation (\ref{eq:ibp}) remains valid 
also for 2dHPL. 

Two-dimensional harmonic polylogarithms can be expressed in
terms of generalised polylogarithms only up to weight 3. At weight 4,
only some special cases relate to  generalised polylogarithms. 
The relations at weight 2 read:
\begin{eqnarray}
  H(0,1-z;y) &=& \Li_2\left(\frac{y}{1-z}\right) \ , \nonumber \\
  H(0,z;y) &=& -\Li_2\left(-\frac{y}{z}\right)  \ , \nonumber \\
  H(1,1-z;y) &=& \frac{1}{2} \ln^2(1-y) - \ln (1-y)\ln (1-z)
                 + \Li_2\left(\frac{z}{1-y}\right) 
                 - \Li_2(z) \ , \nonumber \\
  H(1,z;y) &=& -\ln\left(\frac{1-y}{1+z}\right) 
                \ln\left(\frac{y+z}{z}\right) 
                + \Li_2\left(\frac{z}{1+z}\right)
                - \Li_2\left(\frac{y+z}{1+z}\right)     \ , \nonumber \\
  H(1-z,0;y) &=& -\ln y \ln\left(1-\frac{y}{1-z}\right)
                   - \Li_2\left(\frac{y}{1-z}\right) \ , \nonumber \\
  H(1-z,1;y) &=& -\frac{1}{2} \ln^2(1-y) + \ln (1-y-z)\ln (1-y)
                 - \Li_2\left(\frac{z}{1-y}\right) 
                 + \Li_2(z) \ , \nonumber \\
  H(z,0;y) &=& \ln y \ln\left(\frac{y+z}{z}\right) 
               + \Li_2\left(-\frac{y}{z}\right)  \ , \nonumber \\
  H(z,1;y) &=& -\ln(1+z) \ln\left(\frac{y+z}{z}\right) 
               - \Li_2\left(\frac{z}{1+z}\right)
               + \Li_2\left(\frac{y+z}{1+z}\right)     \ , \nonumber \\
  H(1-z,z;y) &=& -\ln(1-y-z)\ln\left(\frac{y+z}{z}\right)
                 + \Li_2(z) - \Li_2(y+z)  \ , \nonumber \\
  H(z,1-z;y) &=& \ln(1-z) \ln\left(\frac{y+z}{z}\right)
                 - \Li_2(z) + \Li_2(y+z)   \ , \nonumber \\
  H(1-z,1-z;y) &=& \frac{1}{2!} \ln^2\left(1-\frac{y}{1-z}\right) 
                        \ , \nonumber \\
  H(z,z;y) &=& \frac{1}{2!} \ln^2\left(\frac{y+z}{z}\right) \ . 
\end{eqnarray}
The $z$-independent HPL $H(0,1;y)$ and $H(1,0;y)$ are also part of the 
full set of 2dHPL at weight 2.

At weight 3, a total of 64 2dHPL exist. Eight of these involve only $(0,1)$ 
in the vector of parameters and are therefore independent of $z$. 
The remaining 56 can all be expressed as 
combinations of generalised polylogarithms up to weight 3. 
Solving the product and integration-by-parts identities, 
the 2dHPL at weight 3 can be expressed by a minimal set of 20 functions.
This minimal set contains 
the two $z$-independent functions $H(0,0,1;y)$ and $H(0,1,1;y)$ and 
18 genuinely two-dimensional functions:
\begin{eqnarray}
  H(0,0,1-z;y) & = & \Li_3\left(\frac{y}{1-z}\right) \ , \nonumber \\
  H(0,0,z;y)   & = & - \Li_3\left( - \frac{y}{z}\right) \ , \nonumber \\
  H(0,1,1-z;y) & = &  \Li_3\left(\frac{-y}{1-y-z}\right)
                     +\Li_3\left(\frac{-z}{1-y-z}\right)
                     -\Li_3\left(\frac{-yz}{1-y-z}\right)
                     + \Li_3(y) \nonumber \\ 
                &&     
                     - \Li_3\left(\frac{-z}{1-z}\right)
                     - \ln\left(1-\frac{y}{1-z}\right) \Li_2(z)
                     - \ln\left(1-\frac{y}{1-z}\right) \Li_2(y)
                      \nonumber \\ 
                &&     
                     - \frac{1}{6} \ln^3(1-y-z) + \frac{1}{6} \ln^3(1-z)
                  \ , \nonumber \\
  H(0,1,z;y) & = &   \Li_3 \left(\frac{y(1+z)}{y+z} \right)
                   - \Li_3 \left(\frac{y}{y+z} \right)
                   - \Li_3 \left(\frac{y+z}{1+z} \right)
                   +  \Li_3 \left(\frac{z}{1+z} \right)
                   - \Li_3(y) \nonumber \\ 
                &&     
                   + \ln\left( \frac{y+z}{z} \right) 
                       \Li_2\left(\frac{z}{1+z} \right)
                   + \ln\left( \frac{y+z}{z} \right) \Li_2 (y) 
                   - \ln z \ln (1+z) 
                             \ln\left( \frac{y+z}{z} \right) \nonumber \\ 
                &&     
                   + \frac{1}{2} \ln (1+z) \ln^2 (y+z)
                   - \frac{1}{2} \ln (1+z) \ln^2 z \ , \nonumber \\
  H(0,1-z,1;y) & = & \S_{1,2}(y) 
                     - \Li_3\left(\frac{yz}{(1-y)(1-z)}\right)
                     + \Li_3\left(\frac{y}{1-z}\right)
                     + \Li_3\left(\frac{z}{1-y}\right)\nonumber \\ 
                &&     
                     - \Li_3(y) -\Li_3(z) 
                     - \ln(1-y) \Li_2\left(\frac{y}{1-z}\right)
                     + \ln(1-y) \Li_2(y)
                     + \ln(1-y) \Li_2(z) 
                  \nonumber \\ 
                &&     
                     + \frac{1}{2} \ln (1-z) \ln^2(1-y) \ , \nonumber \\
  H(0,1-z,1-z;y) & = & \S_{1,2} \left(\frac{y}{1-z}\right) \ , \nonumber \\
  H(0,1-z,z;y)  & = &    \Li_3\left( \frac{y}{(1-z)(y+z)} \right) 
                     - \Li_3\left( \frac{y}{1-z} \right) 
                     - \Li_3\left( \frac{y}{y+z} \right)
                     - \Li_3(y+z) + \Li_3(z) \nonumber \\ 
                &&     
                     + \ln\left(\frac{y+z}{z}\right) \Li_2(z) 
                     + \ln\left(\frac{y+z}{z}\right) 
                        \Li_2\left(\frac{y}{1-z}\right)
                     - \frac{1}{2} \ln (1-z)\ln^2\left(\frac{y+z}{z}\right) 
                       \ , \nonumber \\
  H(0,z,1;y) & = & -\S_{1,2} (y) + \Li_3\left(- \frac{y(1+z)}{z(1-y)}\right)
                   - \Li_3\left(- \frac{y}{z}\right) 
                   - \Li_3\left(\frac{1-y}{1+z}\right) 
                   + \Li_3\left(\frac{1}{1+z}\right) \nonumber \\ 
                &&     
                   + \Li_3(y) 
                   + \ln (1-y) \Li_2\left(- \frac{y}{z}\right) 
                   + \ln (1-y) \Li_2\left( \frac{1}{1+z}\right) 
                   - \ln (1-y) \Li_2(y) \nonumber \\ 
                &&     
                   + \frac{1}{2} \ln\left(\frac{1+z}{z}\right) \ln^2 (1-y) 
                   - \frac{1}{6} \ln^3(1-y)
\ , \nonumber \\
  H(0,z,1-z;y) & = & \S_{1,2} \left(\frac{y}{(1-z)(y+z)}\right)
                    -\S_{1,2} \left(\frac{y}{1-z}\right)
                    -\S_{1,2} \left(\frac{y}{y+z}\right)
                    +\S_{1,2} \left(y+z\right) \nonumber \\ 
                &&     
                    - \S_{1,2}(z) 
                    -\Li_3 \left(\frac{y}{(1-z)(y+z)}\right)
                    +\Li_3 \left(\frac{y}{1-z}\right)
                    +\Li_3 \left(\frac{y}{y+z}\right)\nonumber \\ 
                &&     
                    +\ln\left(\frac{(1-z)(y+z)}{z}\right)
                       \left ( \Li_2 (y+z) - \Li_2(z) \right)
                    -\ln\left(\frac{y+z}{z}\right)
                        \Li_2 \left(\frac{y}{1-z}\right)\nonumber \\ 
                &&     
                    + \ln(1-z) \ln^2\left(\frac{y+z}{z}\right)
                    + \frac{1}{2}  \ln^2(1-z) \ln\left(\frac{y+z}{z}\right)
\ , \nonumber \\
  H(0,z,z;y) & = & \S_{1,2} \left(- \frac{y}{z}\right) \ , \nonumber \\
  H(1,1-z,1-z;y)  & = & -\frac{1}{2} \ln\left(\frac{1-y}{z}\right) 
                          \ln^2\left(\frac{1-y-z}{1-z}\right)
                        - \ln\left(\frac{1-y-z}{1-z}\right)
                              \Li_2\left(1- \frac{1-y}{z}\right) \nonumber \\
                  &&
                        - \Li_3\left( 1-\frac{1}{z}\right)
                        + \Li_3\left(1 -\frac{1-y}{z}\right)\ , \nonumber \\
  H(1,1-z,z;y)  & = & 2  \S_{1,2}(y)- 2 \Li_3\left(\frac{z}{y+z}\right) 
                     + \Li_3\left(\frac{z}{(1-y)(y+z)}\right)
                     - \Li_3\left(\frac{z}{1-y}\right) \nonumber \\ 
                &&     
                     + \Li_3\left(\frac{z(1-y)}{y+z}\right)
                     - \Li_3\left(- \frac{1-y}{y+z} \right) 
                     - \Li_3(y+z)
                     + \Li_3\left(-\frac{1}{z}\right)
                     + 2 \Li_3(z)  \nonumber \\ 
                &&     
                     + \ln \left(\frac{y+z}{z}\right) 
                         \Li_2\left(\frac{z}{1-y}\right)
                     + \ln \left(\frac{y+z}{z(1-y)}\right) 
                         \Li_2\left(\frac{1}{1+z}\right)
                     - \ln (1-y) \Li_2(z) 
                      \nonumber \\ 
                &&     
                     + 2 \ln (1-y) \Li_2(y) 
                     + \ln y \ln^2(1-y)                    
                     - \frac{1}{2} \ln z \ln^2\left(\frac{1-y}{1+z}\right)
                       \nonumber\\ 
                &&   
                     -\ln z\ln(1+z)\ln\left(\frac{y+z}{z}\right)  
                     - \frac{1}{2} \ln(1-y)\ln^2(y+z)
                     +  \frac{1}{2} \ln\left(\frac{y+z}{1-y}\right) \ln^2(1+z)
                       \nonumber\\ 
                &&   
                     + \frac{1}{6} \ln^3 (y+z) 
                    -  \frac{1}{6} \ln^3 z
\ , \nonumber \\
  H(1,z,1-z;y)  & = & \S_{1,2} \left(\frac{z(y+z)}{1-y}\right)
                     -\S_{1,2} \left(\frac{z}{1-y}\right)
                     -\S_{1,2} \left(y+z\right) - \S_{1,2} (z^2) 
                     + 2\S_{1,2}(z) \nonumber \\ 
                &&     
                     + \Li_3\left(\frac{1-y}{1+z}\right)
                     - \Li_3\left(\frac{1}{1+z}\right)
                     + \ln (1+z) \Li_2 \left(\frac{z}{1-y}\right) 
                     - \ln (1+z) \Li_2 \left(z\right) \nonumber \\ 
                &&     
                     + \ln\left(\frac{1-z^2}{1-y}\right) 
                           \Li_2\left(\frac{1-y}{1+z}\right) 
                     - \ln\left(1-z^2\right) \Li_2\left(\frac{1}{1+z}\right) 
                     + \ln \left(\frac{1+z}{1-y}\right) \Li_2(y+z)\nonumber \\ 
                &&     
                     - \ln \left(\frac{1+z}{1-y}\right) \Li_2(z)
                     + \ln z \ln (1-y)\ln(1-z) 
                     + \frac{1}{2} \ln^2(1-y) \ln\left(\frac{1+z}{y+z}\right)
\nonumber \\ 
                &&     
                     + \ln (1-y) \ln(1+z) \ln\left(\frac{y+z}{1-z}\right)
                     - \frac{1}{2} \ln (1-y) \ln^2 (1+z)\nonumber \\ 
                &&     
                     -\frac{1}{2} \ln^2(1+z) \ln\left(\frac{y+z}{z}\right)
 \ , \nonumber \\
  H(1,z,z;y)  & = & -\frac{1}{2} \ln\left(\frac{1-y}{1+z}\right)
                      \ln^2\left(\frac{y+z}{z}\right) 
                   -\ln\left(\frac{y+z}{z}\right) 
                      \Li_2\left(\frac{y+z}{1+z}\right) 
                   - \Li_3\left(\frac{z}{1+z}\right)\nonumber \\
                  &&
                   + \Li_3\left(\frac{y+z}{1+z}\right)\ , \nonumber \\
  H(1-z,1,1;y) & = &  \frac{1}{2} \ln^2z \ln(1-z) 
                     - \frac{1}{2} \ln^2z \ln(1-y-z)
                     + \ln z \Li_2\left( 1- \frac{1-y}{z} \right) \nonumber \\
                  &&
                     - \ln z \Li_2\left( 1- \frac{1}{z} \right)
                     - \S_{1,2}\left( 1- \frac{1-y}{z} \right) 
                     + \S_{1,2}\left( 1- \frac{1}{z} \right)
\ , \nonumber \\
  H(1-z,z,z;y) & = & -\frac{1}{2}\ln(1-y-z)\ln^2\left(\frac{y+z}{z}\right)  
                         -\ln\left(\frac{y+z}{z}\right)\Li_2(y+z) 
                         -\Li_3(z) + \Li_3(y+z) \ , \nonumber \\
  H(z,1,1;y) & = & \frac{1}{2} \ln^2(1-y) \ln\left(\frac{y+z}{1+z}\right)
                   + \ln(1-y)\Li_2\left(\frac{1-y}{1+z}\right)
                   + \Li_3\left(\frac{1}{1+z}\right) \nonumber \\
                  &&
                   - \Li_3\left(\frac{1-y}{1+z}\right) \ , \nonumber \\
  H(z,1-z,1-z;y) & = & \frac{1}{2}\ln(y+z)\ln^2\left(\frac{1-y-z}{1-z}\right) 
                       + \ln\left(\frac{1-y-z}{1-z}\right)\Li_2(1-y-z)
                       + \Li_3(1-z)\nonumber \\
                  && - \Li_3(1-y-z)\ .
\end{eqnarray}
To illustrate the relation of arbitrary 2dHPL at weight 3 to the respective 
minimal set, let us consider $H(1-z,1,0;y)$, which is in fact the only 
2dHPL appearing in the divergent $1/\e$-part of the planar master integrals.
By writing out the integration-by-parts identity (\ref{eq:ibp}) for 
$H(0,1,1-z;y)$,
we immediately arrive at:
\begin{equation}
H(1-z,1,0;y) = H(1-z;y) H(1,0;y)- H(1,1-z;y) H(0;y) 
              + H(0,1,1-z;y)\; .
\end{equation}

The full basis of 2dHPL at weight 4 contains 256 functions, of which 
16 are independent of $z$. The minimal set at this weight consists of 
60 functions, 3 of them independent of $z$. Only particular combinations 
of 2dHPL at weight 4 can be expressed in terms of generalised 
polylogarithms. In general,
the 2dHPL at weight 4 can 
always be written as one-dimensional integrals over the 2dHPL of weight 3 
given above:
\begin{equation}
H(n,\vec{m}_3;y) = \int_0^y \d y^{\prime} f(n,y^{\prime}) 
H(\vec{m}_3,y^{\prime})\;. 
\end{equation}

\subsubsection{Limiting cases}
\label{sec:limit}
If $y$ or $z$ are fixed to particular numerical values, or if they are 
related to each other, 2dHPL reduce to ordinary HPL. We have to derive
reduction formulae for several cases, which are relevant to 
match the boundary conditions of the differential equations we 
study in this paper. 

In $z=0$ and $z=1$, the reduction from 2dHPL to HPL is a mere substitution 
on the components of $\vec{m}_w$, combined with an adjustment of the 
overall sign of the HPL under consideration. The corresponding rules can 
be derived by writing out the 2dHPL as a multiple integral by repeated 
application of (\ref{eq:inth}):
\begin{equation}
H(\vec{m}_w;y) 
= \int_0^y \d t_1 f(m_1;t_1) \int_0^{t_1} \d t_2 f(m_2;t_2)\ldots
\int_0^{t_{w-1}} \d t_w f(m_w;t_w)\; , 
\label{eq:zlimone}
\end{equation}
valid for $m_w\neq 0$, and
\begin{equation}
H(\vec{m}_w;y) 
= \int_0^y \d t_1 f(m_1;t_1) \int_0^{t_1} \d t_2 f(m_2;t_2)\ldots
\int_0^{t_{v-1}} \d t_v f(m_v;t_v) H(\vec{0}_{w-v};t_v)\; , 
\label{eq:zlimtwo}
\end{equation}
valid for $\vec{m}_w = (\vec{m}_v,\vec{0}_{w-v})$ with $m_v\neq 0$.

If $m_w\neq z$ (respectively $m_v \neq z$ in the second case), 
$H(\vec{m}_w;y)$ remains finite in the limit $z=0$. This limit is
obtained by replacing $f(z;t_i) \to f(0;t_i)$ and 
$f(1-z;t_i) \to f(1;t_i)$ in (\ref{eq:zlimone},\ref{eq:zlimtwo}) and 
subsequent integration. For 
$m_w= z$ (respectively $m_v= z$), $H(\vec{m}_w;y)$ diverges 
proportional to  $\ln z$ 
in $z\to 0$. This divergence can be made explicit by the use of the
HPL-algebra, such that $H(\vec{m}_w;y)$ can be written as 
linear combination  
of logarithmically divergent and finite terms in $z\to 0$.

If $m_w\neq 1-z$, (respectively $m_v \neq 1-z$),
 $H(\vec{m}_w;y)$ remains finite in the limit $z=1$, it develops 
divergent terms proportional to $\ln (1-z)$ otherwise. The limit 
$z=1$ is obtained by replacing $f(z;t_i) \to f(-1;t_i)$ and 
$f(1-z;t_i) \to -f(0;t_i)$  in (\ref{eq:zlimone},\ref{eq:zlimtwo}) and 
subsequent integration.

To reduce 2dHPL in $y=1-z$, $y=-z$ or $y=1$ into HPL of argument $z$, 
one uses
\begin{equation}
H(\vec{m}(z);y(z)) = H(\vec{m}(z=0);y(z=0)) + \int_0^z \d z^{\prime}
                     \frac{\d}{\d z^{\prime}} H(\vec{m}(z^{\prime});y(z^{\prime}))\; ,
\end{equation}
where the boundary $z=0$ can be replaced by $z=1$ if 
$H(\vec{m}(z=0);y(z=0))$ is divergent. The derivative 
$\frac{\d}{\d z} H(\vec{m}(z);y(z))$ is then carried out on the 
multiple integral representation of $H(\vec{m}(z);y(z))$ 
(\ref{eq:zlimone},\ref{eq:zlimtwo}). The resulting $[f(m_i;t_i)]^2$ 
are reduced to single powers by repeated integration by parts and
partial fractioning. The resulting multiple integral can then be
identified as multiple integral representation of 
an ordinary HPL $H(\vec{m};z)$. As an example, we evaluate 
$ H(1,1-z,y) $ in $y=1-z$ : 
\begin{eqnarray}
H(1,1-z;1-z) & = & H(1,0;0) + \int_1^z \d z^{\prime} \frac{\d }{\d z^{\prime}} 
                          H(1,1-z^{\prime};1-z^{\prime})
                 \nonumber \\
&=& \int_1^z \d z^{\prime} \frac{\d }{\d z^{\prime}} 
                   \int_0^{1-z^{\prime}} \frac{\d t_1}{1-t_1}
                                   \int_0^{t_1} \frac{\d t_2}{1-t_2-z^{\prime}}
                 \nonumber \\
&=& \int_1^z \d z^{\prime} \left[ -\frac{1}{z^{\prime}} 
                              \int_0^{1-z^{\prime}} \frac{\d
                           t_2}{1-t_2-z^{\prime}}
                         + \int_0^{1-z^{\prime}} \frac{\d t_1}{1-t_1} 
                           \int_0^{t_1} 
                              \frac{\d t_2}{(1-t_2-z^{\prime})^2}\right]
                 \nonumber \\
&=& \int_1^z \d z^{\prime} \left(- \frac{1}{1-z^{\prime}}
                               -\frac{1}{z^{\prime}}\right) 
                     \int_0^{1-z^{\prime}} \frac{\d t_1}{1-t_1} 
                 \nonumber \\
&=& \int_1^z \d z^{\prime} \left(- \frac{1}{1-z^{\prime}}
                            -\frac{1}{z^{\prime}}\right)  H(1;1-z^{\prime})
                 \nonumber \\
&=& \int_1^z \d z^{\prime} \left(+ \frac{1}{1-z^{\prime}}
                        +\frac{1}{z^{\prime}}\right)  H(0;z^{\prime})
                 \nonumber \\
&=& H(0,0;z) + H(1,0;z) - H(1,0;1) \; . 
\end{eqnarray}
To derive the transformation formulae in $y=1-z$, $y=-z$ or $y=1$ for
2dHPL up to weight 4, we have programmed the underlying algorithm in
FORM~\cite{form}. It is evident from the above example that the
transformation is carried out in a recursive manner, i.e.\ by using the 
results for 2dHPL of weight $w-1$ in the transformation of 2dHPL 
of weight $w$.

\subsubsection{Interchange of arguments}
To check our results for the master integrals, which were obtained in 
terms of $H(\vec{m}(z);y)$ by solving the $y$ differential equations, 
we inserted them into the corresponding $z$ differential equations.
To carry out this check, we have to convert 
the 2dHPL $H(\vec{m}(z);y)$ into 2dHPL $H(\vec{m}(y);z)$ and 
ordinary HPL $H(\vec{m};y)$. This conversion is made using 
\begin{equation}
H(\vec{m}(z);y) = H(\vec{m}(z=0);y) + \int_0^z \d z^{\prime}
                     \frac{\d}{\d z^{\prime}} H(\vec{m}(z^{\prime});y)\; ,
\end{equation}
where the boundary $z=0$ can be replaced by $z=1$ if 
$H(\vec{m}(z=0);y)$ is divergent.
The $z$-derivative is again carried out on the 
multiple integral representation of $H(\vec{m}(z);y)$ 
(\ref{eq:zlimone},\ref{eq:zlimtwo}). Repeated application of 
integration by parts and partial fractioning does then generate a form
that can be identified as a bilinear combination of 
2dHPL $H(\vec{m}(y);z)$ and ordinary HPL $H(\vec{m};y)$.
As an example, consider 
\begin{eqnarray}
H(1,1-z;y) & = & H(1,1;y) + \int_0^z \d z^{\prime} \frac{\d}{\d z^{\prime}} H(1,1-z^{\prime};y)
                 \nonumber \\
&=& H(1,1;y) + \int_0^z \d z^{\prime} \frac{\d}{\d z^{\prime}} \int_0^y 
                                           \frac{\d t_1}{1-t_1}
                               \int_0^{t_1}\frac{\d t_2}{1-t_2-z^{\prime}} 
                 \nonumber \\
&=& H(1,1;y) + \int_0^z \d z^{\prime} \int_0^y \frac{\d t_1}{1-t_1}
                               \int_0^{t_1}\frac{\d t_2}{(1-t_2-z^{\prime})^2} 
                 \nonumber \\
&=& H(1,1;y) + \int_0^z \d z^{\prime} \int_0^y \d t_1 \left[
                                       \left(- \frac{1}{1-z^{\prime}} 
                                            - \frac{1}{z^{\prime}}\right) 
                                          \frac{1}{1-t_1} 
                         + \frac{1}{z^{\prime}} \frac{1}{1-t_1-z^{\prime}} \right]
                 \nonumber \\
&=& H(1,1;y) + \int_0^z \d z^{\prime} \left[\left(- \frac{1}{1-z^{\prime}} 
            - \frac{1}{z^{\prime}}\right) H(1;y) + \frac{1}{z^{\prime}} H(1-z^{\prime};y)\right]
                 \nonumber \\
&=&  H(1,1;y) - H(1;y) \left[ H(0;z) + H(1;z) \right] 
           + \int_0^z \frac{\d z^{\prime}}{z^{\prime}} 
                   \left[H(1-y;z^{\prime}) + H(1;y) - H(1;z^{\prime}) \right]
                 \nonumber \\
&=& H(1,1;y) + H(0,1-y;z) - H(0,1;z) - H(1;y)H(1;z) \; .
\end{eqnarray}
Interchange formulae have been derived for 2dHPL up to weight 4 by
programming the underlying algorithm in FORM~\cite{form}. Like the
transformation algorithm presented above, the interchange algorithm also 
works recursively by using interchange formulae obtained for lower
weights. 

\subsubsection{Analytic continuation}
The 2dHPL introduced in this appendix are real for the following range of 
arguments: $0 \leq y \leq 1$, $0 \leq z \leq 1-y$. This region of 
$y=\sac/\sabc$ and $z=\sbc/\sabc$ is the kinematical region corresponding 
to a $p_{123}\to p_1+p_2+p_3$ 
particle decay process, as in $e^+e^- \to 3$~Jets. To use the 
results obtained in this paper in the computation of virtual corrections
to other processes, which are related to  $1\to 3$ particle decay 
by crossing, one needs the analytic continuation of 2dHPL into regions 
of $y$ and $z$ outside the above-mentioned area. 

This task does a priori not seem feasible, since the 
causal prescription 
\begin{eqnarray}
\sab &=& (p_1+p_2)^2 + i\delta \nonumber \\
\sac &=& (p_1+p_3)^2 + i\delta \nonumber \\
\sbc &=& (p_2+p_3)^2 + i\delta
\end{eqnarray}
is broken by the 
elimination of
$\sab = \sabc - \sac -\sbc = \sabc(1-y-z)$ 
from all expressions.  The 2dHPL formalism does however allow to 
recover the correct analytic structure of the master integrals.

\begin{figure}[t]
\begin{center}
\begin{picture}(12,9)
\thicklines
\put(0,4){\vector(1,0){12}}
\put(4,0){\vector(0,1){9}}
\thinlines
\put(4,7){\line(1,0){6}}
\put(4,7){\line(1,-1){6}}
\put(7,8){\line(0,-1){7}}
\put(4.8,4.8){(I)}
\put(6,6){(II)}
\put(9,5.5){(III)}
\put(9,3){(IV)}
\put(3.6,8.6){$z$}
\put(11.5,3.5){$y$}
\put(7,0.9){\makebox(0,0)[t]{$z=1$}}
\put(10,0.9){\makebox(0,0)[t]{$z=1-y$}}
\put(10.1,7){\makebox(0,0)[l]{$y=1$}}
\end{picture}
\end{center}
\caption{Kinematic regions for the analytic continuation of the 
2dHPL (example)}
\label{fig:kin}
\end{figure}
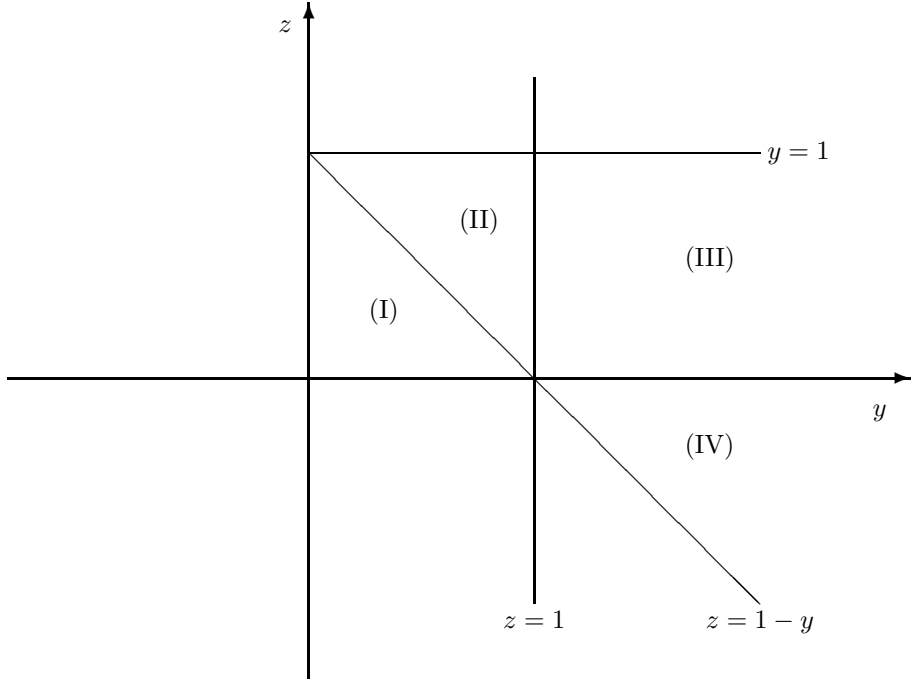
In this subsection, we 
discuss the analytic continuation of 2dHPL to the kinematical 
situation relevant for vector boson plus jet production in hadron collisions:
$2\to 2$ scattering with one final state momentum off shell. The kinematical 
region for this process, with $p_1$ and $p_3$ being the 
incoming momenta, is defined by
\begin{equation}
\sac \geq \sabc \, , 0 \geq \sbc \geq \sabc - \sac \; ,
\end{equation}
corresponding to
\begin{equation}
y \geq 1 \, , 0 \geq z \geq 1-y \; .
\end{equation}
This kinematical region is denoted by (IV) in Figure~\ref{fig:kin}, while 
the region $0 \leq y \leq 1$, $0 \leq z \leq 1-y$ is denoted by (I). 
To perform the analytic continuation of the 2dHPL from region (I) to region 
(IV), we proceed as follows:
\begin{enumerate}
\item Continuation from region (I) to the boundary of regions (II) and (III), 
      defined by $y=1$, $0 \leq z \leq 1$. After separation of the 
      divergent $\ln(1-y)$-terms by using (\ref{eq:alg}), 
      this continuation proceeds recursively as outlined in 
      Section~\ref{sec:limit}.
      By specifying the continuation of 
\begin{equation}
H(1-z;y=1) = -\ln (1-1-z+i\delta) + \ln(1-z) = -\ln z -i\pi + \ln(1-z) = 
- H(0;z) - H(1;z) - i\pi\; ,
\label{eq:aone}
\end{equation}
all imaginary parts of $H(\vec{m}(z);1)$ are specified.
\item Transformation of variables in region (III): $y = 1/(1-u)$ and 
      $z=v/(1-u)$. $H(\vec{m}(z);y)$ are then expressed as 
      linear combination of 2dHPL $H(\vec{m}(v);u)$ and HPL $H(\vec{m};v)$
      by 
\begin{equation}
H(\vec{m}(z);y) = H(\vec{m}(z=v);y=1) + \int_0^u \d u^\prime 
\frac{\d}{\d u^{\prime}} 
H\left(\vec{m}\left(z=\frac{v}{1-u^\prime}\right); 
y = \frac{1}{1-u'}\right)\; ,
\end{equation}
which is well defined since all divergent $\ln(1-y)$-terms were separated 
off beforehand. These terms are continued according to 
\begin{equation}
H(1;y) = -\ln (1-y - i\delta) = -\ln (-u-i\delta) + \ln (1-u) 
= -\ln u + i\pi + \ln (1-u) = -H(0;u)-H(1;u) + i\pi \; .
\label{eq:atwo}
\end{equation}
In region (III), one has $0 \leq u \leq 1$ and $0\leq v \leq 1-u$, 
such that the $H(\vec{m}(v);u)$ and $H(\vec{m};v)$ are real. 
\item Interchange of arguments: The $H(\vec{m}(v);u)$ are transformed 
into a linear combination of $H(\vec{m}(u);v)$ and $H(\vec{m};u)$. 
All terms proportional to $\ln v$ are separated off using the 
product identities, such that the remaining terms are finite in the limit
$v =0$, which is the boundary to region (IV).
\item Transformation of variables in region (IV): $u = 1-\tau$ 
and $v = -\xi$. $H(\vec{m}(u);v)$  are then expressed as linear 
combination of 2dHPL $H(\vec{m}(\tau);\xi)$ and HPL $H(\vec{m};\tau)$
by
\begin{equation}
H(\vec{m}(u);v) = H(\vec{m}(u);v=0) + \int_0^\xi \d \xi^\prime
\frac{\d}{\d \xi^{\prime}}  
H\left(\vec{m}\left(u=1-\tau\right); 
v = -\xi \right)\; ,
\end{equation}
which is again well defined. The remaining $\ln v$-terms are continued 
according to:
\begin{equation}
H(0;v) = \ln \left(\frac{z+i\delta}{y}\right) = \ln(-\xi+i\delta) 
= \ln \xi  + i\pi   = H(0;\xi) + i \pi\;.
\label{eq:athree}
\end{equation}
The $H(\vec{m};u=1-\tau)$ are transformed into $H(\vec{m};\tau)$ by using the 
standard HPL formulae of~\cite{hpl}. 
\end{enumerate}

The newly introduced variables fulfil
\begin{equation}
0 \leq \tau=\frac{\sabc}{\sac} \leq 1\ , \qquad \mbox{and} 
\qquad 0 \leq \xi = \frac{-\sbc}{\sac} \leq 1-\tau\; ,
\end{equation}
which is the region where the 2dHPL
$H(\vec{m}(\tau);\xi)$ and the HPL $H(\vec{m};\tau)$
are real. The imaginary parts have been separated off, and appear explicitly. 

With this example, we have demonstrated how the analytic continuation of 
2dHPL can be carried out in practice. The main point to notice is that the 
analytic continuation formulae (\ref{eq:aone}), (\ref{eq:atwo}) 
and (\ref{eq:athree}) appear to 
be mutually inconsistent as far as the assignment of imaginary parts to the 
variables $y$ and $z$ is concerned. This apparent inconsistency is however 
only an artifact of the identification $\sab = \sabc(1-y-z)$. Considering 
the HPL and 2dHPL at weight 1 as the basic objects 
for analytic continuation (and not the variables $y$ and $z$)
this inconsistency is no longer present. It is 
sufficient to specify the analytic continuation at weight 1, the 
continuation for higher 
weights is obtained recursively from the lower weights.

\end{appendix}

\end{document}